\documentclass[showpacs,preprintnumbers,amsmath,amssymb,APSl,prd,nofootinbib,superscriptaddress, twocolumn]{revtex4-1}

\usepackage{dcolumn}% Align table columns on decimal point
\usepackage{bm}% bold math
\usepackage{ifpdf}
\usepackage{hyperref}
\usepackage{bm}
\usepackage{xcolor,color,graphicx,graphics}
\usepackage[spanish,english]{babel}%, portuguese
\usepackage[latin1]{inputenc}
\usepackage[OT1]{fontenc}
\usepackage{latexsym,amssymb,amsmath,amsfonts}
\usepackage{makeidx}
\usepackage{epsfig,subfigure}
\usepackage{natbib}
\usepackage{epstopdf}
\usepackage{mathrsfs}
\hypersetup{colorlinks=true, linkcolor=blue, citecolor=green}
\usepackage{enumerate}
\usepackage{xcolor}
 \usepackage{multirow}
\definecolor{red}{rgb}{1,0,0}

\def\+{^\dagger}

\def\<{\leftarrow}
\def\>{\rightarrow}

\def\({\left(}
\def\){\right)}

\newcommand{\bi}{\begin{itemize}} 				\newcommand{\ei}{\end{itemize}}
\newcommand{\benu}{\begin{enumerate}} 		\newcommand{\enu}{\end{enumerate}}
\newcommand{\bd}{\begin{dinglist}{0}}     \newcommand{\ed}{\end{dinglist}}
\newcommand{\bfig}{\begin{figure}[htbp]}  \newcommand{\efig}{\end{figure}}
        			
\newcommand{\bc}{\begin{center}} 				  \newcommand{\ec}{\end{center}}
\newcommand{\be}{\begin{equation}} 				\newcommand{\ee}{\end{equation}}
\newcommand{\bsub}{\begin{subequations}}  \newcommand{\esub}{\end{subequations}}
\newcommand{\ben}{\begin{eqnarray}} 			\newcommand{\een}{\end{eqnarray}}
\newcommand{\ba}[1]{\begin{array}{#1}} 		\newcommand{\ea}{\end{array}}
\newcommand{\bea}{\begin{equation}\begin{array}{rcl}}
\newcommand{\eea}{\end{array}\end{equation}}

\begin{document}
\title{Shadows and optical appearance of black bounces illuminated by a thin accretion disk}

\author{Merce Guerrero} \email{merguerr@ucm.es}
\affiliation{Departamento de F\'isica Te\'orica and IPARCOS,
	Universidad Complutense de Madrid, E-28040 Madrid, Spain}
	\author{Gonzalo J. Olmo} \email{gonzalo.olmo@uv.es}
\affiliation{Departamento de F\'{i}sica Te\'{o}rica and IFIC, Centro Mixto Universidad de Valencia - CSIC.
Universidad de Valencia, Burjassot-46100, Valencia, Spain}
\affiliation{Departamento de F\'isica, Universidade Federal da
Para\'\i ba, 58051-900 Jo\~ao Pessoa, Para\'\i ba, Brazil}
\author{Diego Rubiera-Garcia} \email{drubiera@ucm.es}
\affiliation{Departamento de F\'isica Te\'orica and IPARCOS,
	Universidad Complutense de Madrid, E-28040 Madrid, Spain}
	\author{Diego S\'aez-Chill\'on G\'omez} \email{diego.saez@uva.es}
\affiliation{Department of Theoretical Physics, Atomic and Optics, Campus Miguel Delibes, \\
University of Valladolid UVA, Paseo Bel\'en, 7, 47011 - Valladolid, Spain}

\date{\today}
\begin{abstract}
We study the light rings and shadows of an uniparametric family of spherically symmetric  geometries interpolating between the Schwarzschild solution, a regular black hole, and a traversable wormhole, and dubbed as black bounces, all of them sharing the same critical impact parameter. We consider the ray-tracing method in order to study the impact parameter regions corresponding to the direct, lensed, and photon ring emission, finding a broadening of all these regions for black bounce solutions as compared to the Schwarzschild one. Using this, we determine the optical appearance of black bounces when illuminated by three standard toy models of optically and geometrically thin accretion disks viewed in face-on orientation.

\end{abstract}

\maketitle

\section{Introduction}

The detection in 2019 by the Einstein Horizon Telescope (EHT) of the accretion flow around the supermassive object at the center of the M87 galaxy \cite{Akiyama:2019cqa} has triggered the beginning of a new era in the analysis of electromagnetic phenomena around compact objects and on testing General Relativity (GR) itself. The image released by the EHT shows a bright ring-shaped lump of radiation surrounding a black central region of an estimated $6.5$ billion solar masses. The canonical interpretation of this phenomenon calls for the deviation of light rays in the gravitational field of an object having a photon sphere (a critical unstable curve) when illuminated by an accretion disk \cite{Falcke:1999pj}. The inner region bounded by this critical curve is commonly known as the black hole shadow \cite{Perlick:2021aok}.

The implications of this discovery are far reaching. Not only does it allow us to test the background geometry using gravitational light deflection (lensing), but also constitute a test of the Kerr hypothesis on the nature of astrophysical black holes as compared to its many competitors \cite{Cardoso:2019rvt}. This is so because  gravitational lensing near critical curves involves testing gravitational effects happening in a regime which is inaccessible to weak-field limit tests \cite{Psaltis:2020lvx} and, as such, it allows plenty of room for alternative compact objects to represent the shadow caster. On the other hand, despite the fact that the main equations governing gravitational lensing are known since a long time ago \cite{Bardeen} (for a in-depth analysis of such equations in the strong-field regime see the paper by Bozza \cite{Bozza:2002zj}) and many non-Kerr black hole shadows beyond GR have been studied in the literature \cite{Johannsen:2010ru,Atamurotov:2013sca,Cunha:2015yba,Abdujabbarov:2016hnw,Held:2019xde,Kumar:2020owy,Xavier:2020egv,Wei:2020ght,Herdeiro:2021lwl,Devi:2021ctm,Hou:2021okc}, only very recently we have come to fully appreciate the great richness of the physics of accretion disks around black holes regarding its impact on the silhouettes of the latter \cite{Narayan:2019imo,Cunha:2019hzj,Boero:2021afh}.

A compact object having a critical curve and illuminated by an accretion disk may yield a complex pattern of contributions to the total luminosity driven by several light ray trajectories. Technically, the critical curve is defined as the light ray received by the observer that, when traced backwards, would have approached asymptotically a bound photon orbit. For a Schwarzschild black hole this determines a critical impact parameter $b_c =3\sqrt{3}M\approx 5.197M$. However,  while the impact parameter region (number of orbits around the critical curve) depends only on the background geometry, the optical appearance of the object is not only a function of it, but also of the geometry and physical properties of the illuminating accretion flow \cite{Gralla:2019xty}. For instance, for an optically and geometrically thin accretion disk the total luminosity is largely dominated by the direct emission (light rays deflected less than $90$ degrees) and, to lower extent, by the lensing ring (light rays that intersect the equatorial plane just twice), with the contribution of the critical curve to it being almost negligible. This is why the authors of \cite{Gralla:2019xty} proposed to call the inner region of this direct emission region (for  Schwarzschild this is $b \lesssim 6.17M$) the black hole shadow, instead of the one associated to the inner region of the critical curve $b=b_c$. Whether this interplay between gravitational theory and the modelling of the accretion disk allows to discriminate GR black holes from alternative compact objects via electromagnetic illumination,  - likewise distinguishing horses from unicorns via their respective shadow - is nowadays an exciting area of research \cite{Glampedakis:2021oie,Junior:2021atr,Chael:2021rjo}.

Once the tabletop is set, the community has launched to the search for shadows cast by different compact objects  when illuminated by different types of accretion disks \cite{Zeng:2020dco,Zeng:2020vsj,Qin:2020xzu,Lima:2020auu,He:2021htq,Peng:2020wun,Gan:2021pwu,Li:2021riw,1865010,Eichhorn:2021iwq,Eichhorn:2021etc,Shaikh:2021cvl} in order to compare them with the GR (Kerr) expectations. Nonetheless, given the many ingredients involved in the analysis of this problem - the underlying background geometry, the assumptions on the symmetries of the problem, the geometrical, optical, and emission aspects of the modeling of the accretions disk, etc -, it is useful to consider some simplifying assumptions in order to investigate prospective smoking guns of new Physics. In this sense, the assumption of spherical symmetry, though seemingly too restrictive given the fact that real astrophysical black holes do rotate, turns out to be a good approximation since the size and shape of the shadow, as seen by an asymptotic observer, depends very weekly on the spin of the black hole in combination with the inclination with respect to the line of sight, with deviations from circularity lying within $\sim 7\%$ for ultra-fast spinning black holes \cite{Psaltis:2018xkc}.

The main aim of this work is to study the optical appearances and shadows of an uniparametric spherically symmetric family of extensions of the Schwarzschild space-time recently introduced in \cite{Simpson:2018tsi} and dubbed as {\it black bounces}, which have attracted quite some attention in the community \cite{Huang:2019arj,Churilova:2019cyt,Lobo:2020kxn,Nascimento:2020ime,Lobo:2020ffi,Tsukamoto:2020bjm,Zhou:2020zys,Mazza:2021rgq,Shaikh:2021yux,Cheng:2021hoc,Islam:2021ful,Fran:2021pyi,Bronnikov:2021liv,Tsukamoto:2021caq,Zeng:2021dlj}. Despite its simple mathematical structure, its interest lies in the following: i) it smoothly interpolates between the Schwarzschild space-time, a family of regular black hole solutions, and a family of traversable wormhole solutions; ii) it has the same critical parameter as in the Schwarzschild solution; iii) it removes the presence of space-time singularities, iv) it has not Cauchy horizons, thus avoiding their associated instability issues \cite{Poisson:1989zz}; v) they can be taken as parameterized deviations from the Schwarzschild solution in a theory-agnostic way (for an example where solutions of this type arise as solutions to modified gravity equations, see \cite{Olmo:2013gqa,Olmo:2015bya,Bejarano:2017fgz}). Since black holes and traversable wormholes are conceptually and operationally two different types of objects, the black bounce geometry allows one to study the light rings and shadows cast by each such object and compare them to that of the Schwarzschild solution. To this end, in this paper we shall characterize the impact parameter regions for each direct/lensed/photon ring trajectories using the ray-tracing method, and moreover consider three standard toy models of geometrically and optically thin accretion disks with different emission profiles in order to find the corresponding optical appearances as compared to the Schwarzschild solution.

This paper is organized as follows: in Sec. \ref{sec:II} we describe the main aspects of the black bounce geometries and  discuss their geodesic motion equations and associated effective potential. In Sec. \ref{sec:III} we use the ray-tracing method in order to study the impact parameter region for the three types of emission (direct/lensed/photon ring) for the different regions of interest of the black bounce parameter. In Sec. \ref{sec:IV} we use the three toy models for the emission profile of the accretion disk in order to study the observational appearance of some samples of black bounces corresponding to the regular black hole and traversable wormhole geometries. Finally in Sec. \ref{sec:V} we summarize our main findings, discuss the limitations of our approach as well as future prospects.

\section{Black bounces} \label{sec:II}

\subsection{Geometry and horizons}

Let us start by considering a static, spherically symmetric solution of the form
\begin{equation} \label{eq:sss}
ds^2=-A(x)dt^2+B(x)dx^2+r^2(x)d\Omega^2 \ ,
\end{equation}
where the radial coordinate $x$ spans the entire real line, $x \in(-\infty,+\infty)$, while $d\Omega^2=d\theta^2 + \sin^2 \theta d\phi^2$ is the line element on the two spheres. The areal radius is measured by $S=4\pi r^2(x)$ and, in bouncing geometries such as in wormhole ones, the radial function $r(x)$ is bounded by $r \geq r_{th}$ in a model-dependent way \cite{VisserBook}. One can note that the above line element can be further simplified to just two free functions by introducing a new radial coordinate $dy^2=B(x)dx^2$, though for the purposes of this paper we shall keep it this form.

By black bounce (BB) we refer to the uniparametric family of solutions given by the line element (\ref{eq:sss}) with \cite{Simpson:2018tsi}
\begin{equation} \label{eq:BBBline}
A(x)=B^{-1}(x)=1-\frac{2M}{r(x)} \hspace{0.1cm}; \hspace{0.1cm} r^2(x)=x^2+a^2 \ ,
\end{equation}
where $a$ is the BB parameter, so in this geometry one has the wormhole throat located at $r_{th}^2=a^2$. The most noticeable feature of such geometries is the bounce (hence its name) in the radial function, in a simple implementation of a wormhole geometry extending the Schwarzschild solution via the replacement $x \to r(x)$, such that in the limit  $a \to 0$ one has $r^2(x) \approx x^2$. Whether the bounce is hidden behind an event horizon or not can be found by looking at the location of the horizons, $g^{xx}=A(x)=0$, which in the present case amounts to
\begin{equation} \label{eq:hor}
x_h^{\pm}=\pm \sqrt{4M^2-a^2} \ ,
\end{equation}
where the $\pm$ signs refer to the location of the horizon on both sides of the throat. From these equations it can be easily seen that the bounce will be hidden by an event horizon if $a<2M$, so in this case one finds a regular black hole (BH) geometry\footnote{Indeed, the bounce allows for the extension of geodesics beyond $x=0$ ($r=0$). For an extended discussion on geodesic completeness restoration mechanisms, see e.g. \cite{Carballo-Rubio:2019fnb}.}, while if $a>2M$ the bounce lies above the would-be horizon and the geometry represents instead a traversable wormhole (WH) solution with its throat located at $x_{th}=0$. Note that in terms of the radial function, the BH solution has its horizon at $r_{h}=2M$, while the WH has its throat at $r_{th}=a>2M$ instead, and no horizon is present.  The case $a=2M$ was argued in \cite{Simpson:2018tsi} to correspond to a non-traversable WH and, for the sake of this paper, we shall use it as a limiting case in the transition BH/WH.

\subsection{Geodesic equations}

\begin{figure}[t!]
\includegraphics[width=8.0cm,height=5.5cm]{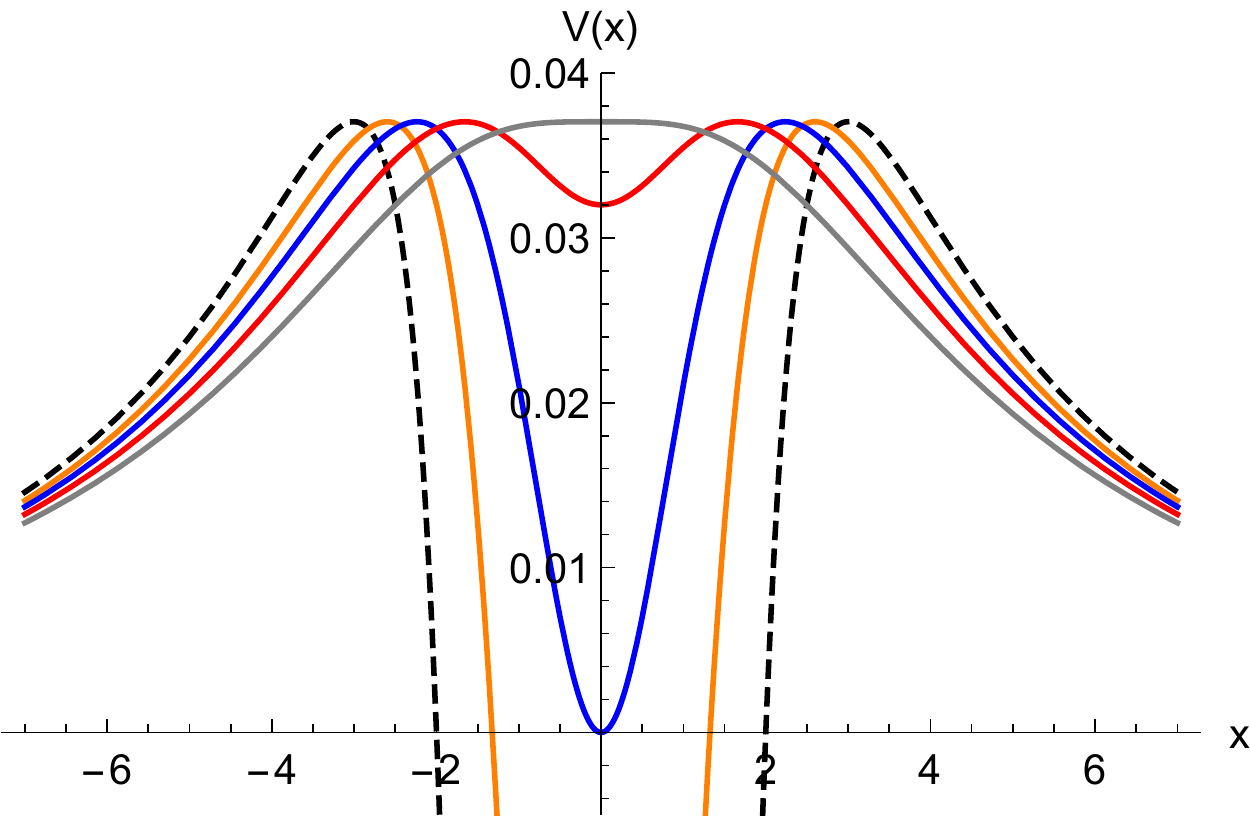}
\caption{The effective potential $V(x)$ in (\ref{eq:Veff}) for the BB solutions with $M=b^2=1$ as a function of $x$ for $a=0$ (dashed black, Schwarzschild solution), $a=3/2$ (orange, BH case), $a=2$ (non-traversable WH, blue), $a=5/2$ (traversable WH, red) and $a=3$ (gray, last photon orbit). Note that only when $a>0$ are both sides of this figure physically connected, since in the $a=0$ case, $r^2 \approx x^2 $ and because $r>0$ then the two regions $x\in(-\infty,0),x\in(0,+\infty)$ are causally disconnected.}
\label{fig:pot}
\end{figure}

A photon travels on a null geodesic, $g^{\mu\nu}k_{\mu}k_{\nu}=0$, with $k^\mu=\dot{x}^{\mu}$ its wave number. In spherical symmetry there are two conserved quantities, namely, the energy per unit mass, $E=-g_{\mu\nu}t^{\mu}k^{\nu}=A\dot{t}$, and the angular momentum per unit mass, $L=g_{\mu\nu}\phi^{\mu}k^{\nu}=r^2  \dot{\phi}$ (dots indicating derivatives with respect to the affine parameter). By spherical symmetry one can assume the motion to take place in the plane $\theta=\pi/2$ without loss of generality, and furthermore by introducing the impact parameter, $b \equiv \tfrac{L}{E}=\tfrac{r^2\dot{\phi}}{A\dot{t}}$, one can cast the geodesic equation for null trajectories as (a re-parametrization of the affine parameter is introduced here to absorb a $L^2$ factor)
\begin{equation} \label{eq:geo}
\dot{x}^2=\frac{1}{b^2}-V(x) \ ,
\end{equation}
which is akin to the equation for a one-dimensional single particle moving in an effective potential of the form
\begin{equation} \label{eq:Veff}
V(x)=\frac{A(x)}{r^2(x)} \ ,
\end{equation}
which is depicted in Fig.\ref{fig:pot} for the BB solution in both the BH and WH cases as compared to the Schwarzschild solution. Let us assume a photon approaching from infinity with a given impact parameter $b$ such that at some value $r_0$ the right-hand side of Eq.(\ref{eq:geo}) vanishes. Such a photon will thus approach to the closest distance $r_0$ before running away back to infinity. The minimum impact factor for which that relation can be satisfied is given by the critical value
\begin{eqnarray} \label{eq:cimp}
b_c^2=\frac{r^2(x_{ps})}{A(x_{ps})} \ ,
\end{eqnarray}
which corresponds to the maximum of the effective potential (\ref{eq:Veff}), i.e., $V_{eff}(x=x_{ps})=\tfrac{1}{b_c^2},V_{eff}'(x=x_{ps})=0, V_{eff}''(x=x_{ps})<0$. At this point $r_{ps}$ it will turn an arbitrarily large number of times around the compact object. However, this orbit is unstable since under a small perturbation the photon will eventually fall into the inner region of the object ($b<b_c$) or escape to asymptotic infinity $(b>b_0$, and therefore this {\it critical curve} (following the notation of \cite{Gralla:2019xty}) defines the unstable null circular orbit. 

In the present BB case, the above conditions define the radius of this critical curve (for which we shall also reserve the word ``photon sphere") as \cite{Bronnikov:2021liv}
\begin{equation} \label{eq:ps}
x_{ps}=\sqrt{9M^2-a^2} \to r_{ps}=3M
\end{equation}
A remarkable property of the BB family of solutions is that, when (\ref{eq:ps}) is introduced in (\ref{eq:cimp}), it yields the critical impact parameter  $b_c=3\sqrt{3}M \approx 5.19615M$ and, therefore, all BB solutions have the same critical impact parameter as the Schwarzschild one. Note that the condition (\ref{eq:ps}) implies that such critical orbits will exist provided that $a<3M$. Therefore the BB configurations relevant for shadows (i.e, having a photon sphere) are naturally split into two families: those with $0<a<2M$ correspond to regular BHs while those with $2M<a<3M$ are traversable WHs, with the $a=2M$ and $a=3M$ acting as limiting cases.

In order to study the optical appearance of a compact object as illuminated by all the light rays passing close by, the geodesic equation (\ref{eq:geo}) must be suitably rewritten in terms of the variation of the azimuthal angle $\phi$ with respect to the radial coordinate, which in the present BB case reads
\begin{equation} \label{eq:geoBB}
\frac{d\phi}{dx}=\mp\frac{b}{r^2(x)\sqrt{1-\tfrac{b^2A(x)}{r^2(x)}}} \ ,
\end{equation}
where the $\mp$ signs refer to ingoing/outgoing trajectories, respectively. The few equations introduced in this section is all the setup we need in order to start with the ray tracing of the BB solutions.

\section{Ray tracing} \label{sec:III}

\begin{figure}[t!]
\includegraphics[width=8.0cm,height=5.5cm]{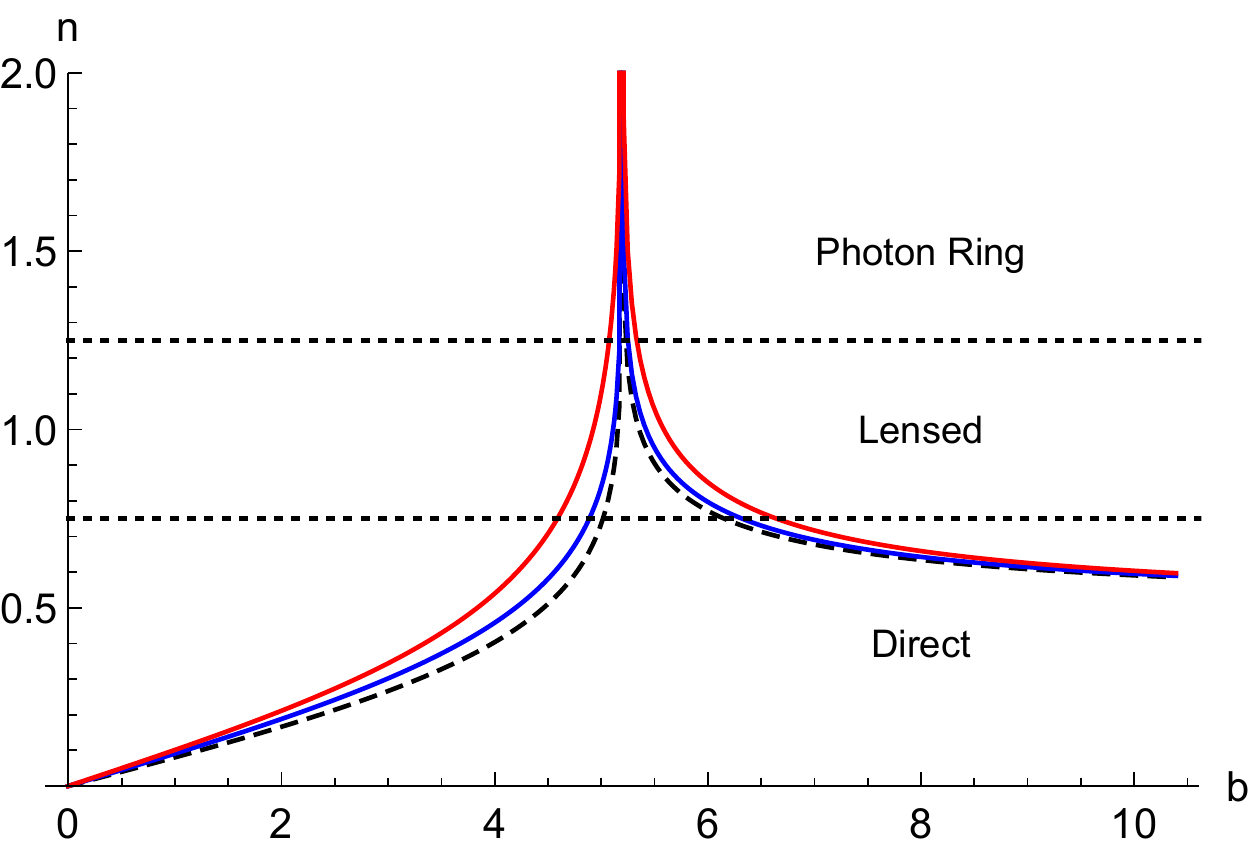}
\caption{The number of orbits, $n \equiv \tfrac{\phi}{2\pi}$, made by a light ray on its trip from its emission source to the observer around the BB solution for a BH with $a=3/2$ (blue) and a traversable WH with $a=5/2$ (red), as compared to the Schwarzschild solution, $a=0$ (dashed black). $n$ diverges at $b=b_c=3\sqrt{3}$ (in units of $M=1$), where it would perform an infinite number of orbits around the BB object.}
\label{fig:noo}
\end{figure}

In the ray-tracing procedure, light rays arriving to the screen of the observer at asymptotic infinity are traced back to the point of the sky they originated from bearing in mind its deflection by the gravitational field of the compact object, which in the BB case is determined by Eq.(\ref{eq:geoBB}). The physical scenario is that of a compact object being illuminated from behind by a planar source which emits isotropically and with uniform brightness. In order to understand the optical appearance of the BB solution by the ray-tracing procedure, we first define the total number of orbits made by a single light ray on its path from its source to the observer as the (normalized) change in the azimuthal angle, that is, $n(b) \equiv \tfrac{\phi}{2\pi}$. This number of orbits will obviously depend on how close the impact parameter is to the critical one and, in addition, on the geometry of the different BB cases. Note that within this setup, light rays in straight motion (i.e. not being deflected at all by the BB solution) have $n=1/2$. In Fig.\ref{fig:noo} we depict the number of orbits for two samples of the BB solutions, representative of the BH and traversable WH families. As expected, the most salient feature of this plot is the narrow spike in both cases at the critical impact parameter $b_c=3\sqrt{3}M$, representing the location of the critical curve where a light ray would have orbited the BB solution an arbitrarily large number of times. For other values of the impact parameter one can see that the larger the BB parameter $a$ is the more orbits for the corresponding BH/WH solutions are found, an effect which is significantly enhanced in the inner region, $b<b_c$, as we move towards the WH solutions.

The next step in our analysis is to integrate the geodesic equation (\ref{eq:geoBB}) for a bunch of light rays spanning the whole region of impact parameter values. For impact parameters $b \geq b_c$ light rays will be deflected at some minimum radius  above the photon sphere one $r=r_0>r_{ps}$, and the corresponding trajectories can be therefore classified according to the number of orbits around the BB solution as follows:
\begin{itemize}
\item Direct emission, for $n<3/4$, corresponding to trajectories that intersect the equatorial plane (on its front side) just once.
\item Lensed emission, for $3/4<n<5/4$, corresponding to trajectories that intersect the equatorial plane twice (on its front and back sides, respectively).
\item Photon ring emission, for $n>5/4$, corresponding to trajectories that intersect the equatorial plane at least three times.
\end{itemize}
It should be noted that for $n>5/4$ one actually finds an infinite sequence of concentric light rings converging to the critical curve $b=b_c$, which are exponentially closer  to each other and thinner as more number of orbits are performed. Therefore, from now on we shall denote by photon ring just the first of such orbits (whose range of impact parameters depends on the BB parameter, see Table \ref{table:I}), and disregard all the others since their contribution to the total luminosity will be negligible \cite{Gralla:2019xty}.

\begin{table*}[t]
\begin{tabular}{| c | c | c | c | c | c | }
\hline
Orbit/BB parameter& $a=0$ & $a=3/2$ & $a=2$ & $a=5/2$ & $a=3$ \\
\hline
Direct 			 & $b>6.17$ & $b>6.32$ & $b>6.46$ & $b>6.64$ & $b>6.88$  \\
\hline
Lensed			 & $5.22<b<6.17$ & $5.25<b<6.32$ & $5.27<b<6.46$ & $5.33<b<6.64$ & $5.44<b<6.88$ \\
\hline
Photon ring 			 & $b_c<b<5.22$ & $b_c<b<5.25$ & $b_c<b<5.27$ & $b_c<b<5.33$ & $b_c<b<5.44$ \\
\hline
Retro-photon	ring 	 & $5.19<b<b_c$	& $5.17<b<b_c$    & $5.04<b<b_c$	  & $5.07<b<b_c$ & $5.13<b<b_c$ \\
\hline
Retro-lensed 	 & $5.02<b<5.18$ 	& $4.89<b<5.17$    & $4.42<b<5.04$ 	  & $4.59<b<5.07$ & $4.83<b<5.13$\\
\hline
Retro-direct		 & $b<5.02$ 	& $b<4.89$    & $b<4.42$ 	  & $b<4.59$ & $b<4.83$ \\
\hline
\end{tabular}
\caption{Range of impact parameters (in units of $M$, and taking only two decimals in order not to overload the text) for different BB cases yielding orbits being deflected above the BB solution photon sphere radius (direct/lensed/photon ring) or emerging from within it (retro-photon ring/retro-lensed/retro-direct). In the BH case  ($a<2$) those-retro-orbits will intersect the event horizon at $x_h=\sqrt{4-a^2}$, while in the WH case ($a \geq 2$)  they will travel all the way down to the wormhole throat $x=0$ ($r_{th}=a>2$). In this plot $b_c=3\sqrt{3} \approx 5.19615$ is the critical impact parameter for all the BB solutions.}
\label{table:I}
\end{table*}

For impact parameters $b<b_c$ the ray tracing yields trajectories that would have emerged from the central region of the object inside the photon sphere (note that having at least one intersection with the disk requires that $n>1/4$). In the BH case, $a<2$, the backtrack of such trajectories intersects with the event horizon, while in the WH case, $a \geq 2$, such trajectories continue their path all the way down to the wormhole throat $x=0$ given the absence of event horizon. For the sake of this paper we find it convenient to also split such $b<b_c$ trajectories emerging out of the internal region of the solutions into three cases depending on the number of orbits:
\begin{itemize}
\item Retro-direct emission, for $n<3/4$.
\item Retro-lensed emission, for $3/4<n<5/4$.
\item Retro-photon ring emission, for $n>5/4$.
\end{itemize}
These retro-orbits could therefore contribute to the luminosity in the observer's screen for impact parameters below the critical one, $b<b_c$, as we shall see in Sec. \ref{sec:IV}.

\begin{figure*}[t]
\centering
\includegraphics[width=5.9cm,height=5.9cm]{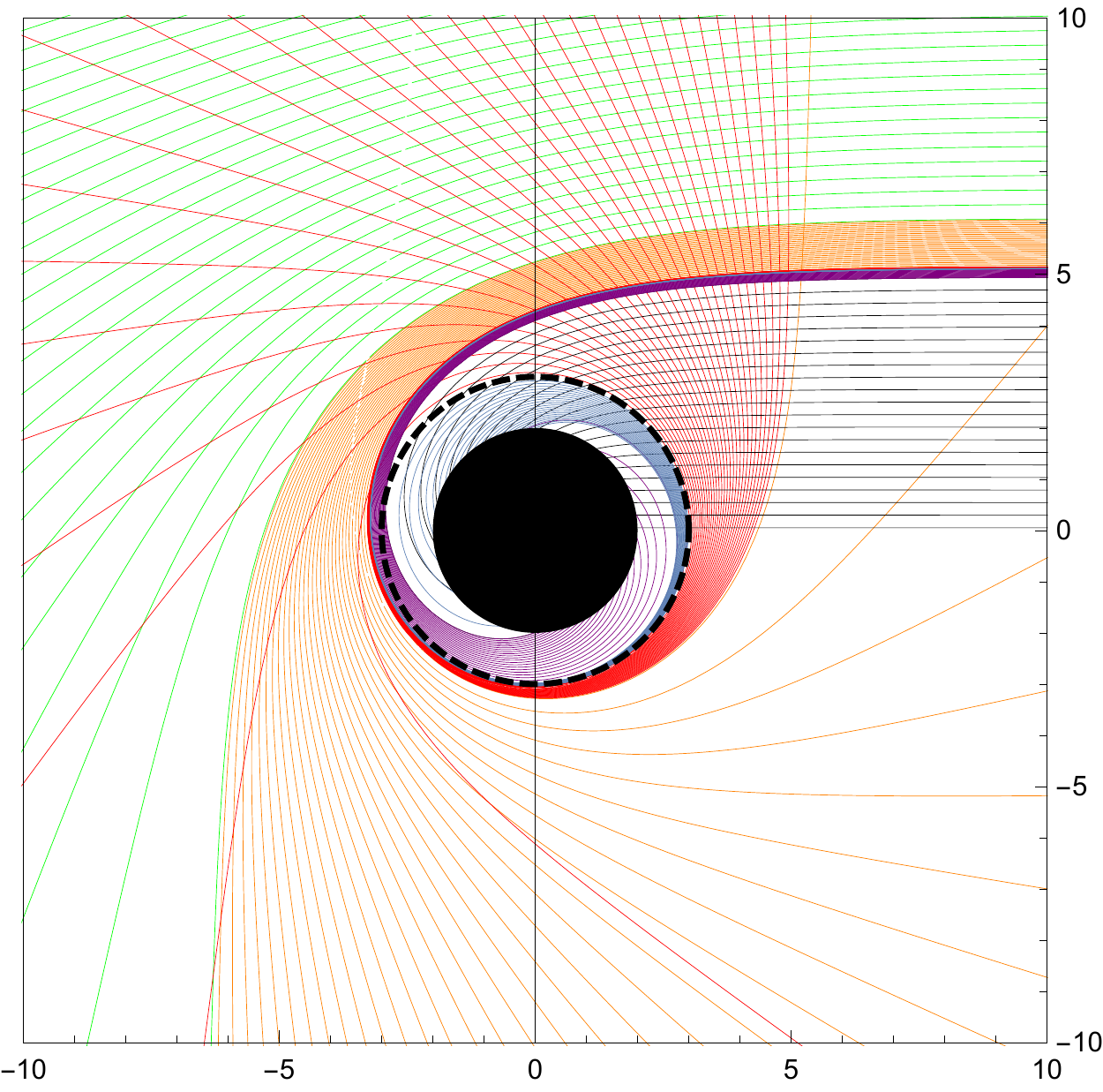}
\includegraphics[width=5.9cm,height=5.9cm]{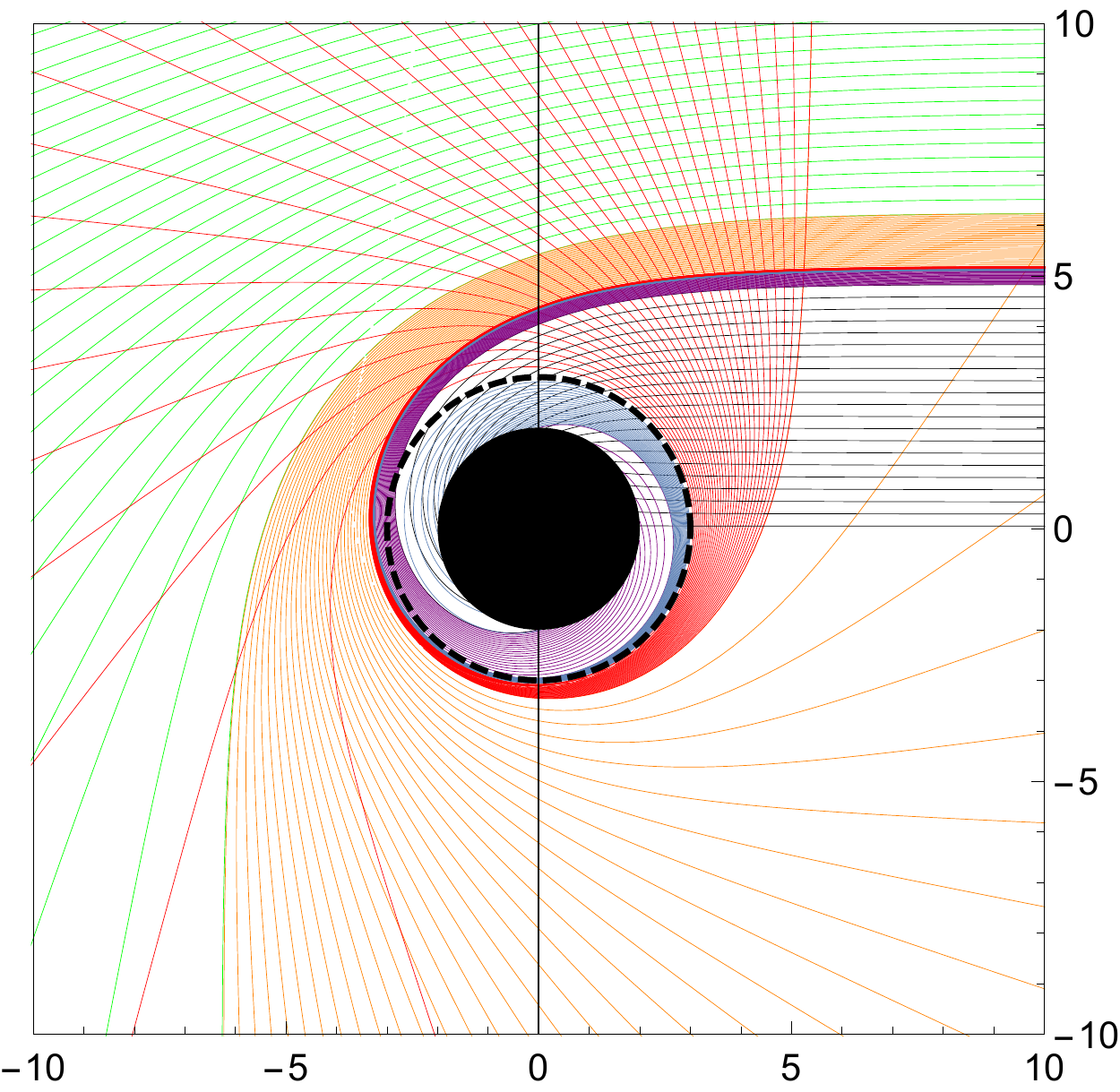}
\includegraphics[width=5.9cm,height=5.9cm]{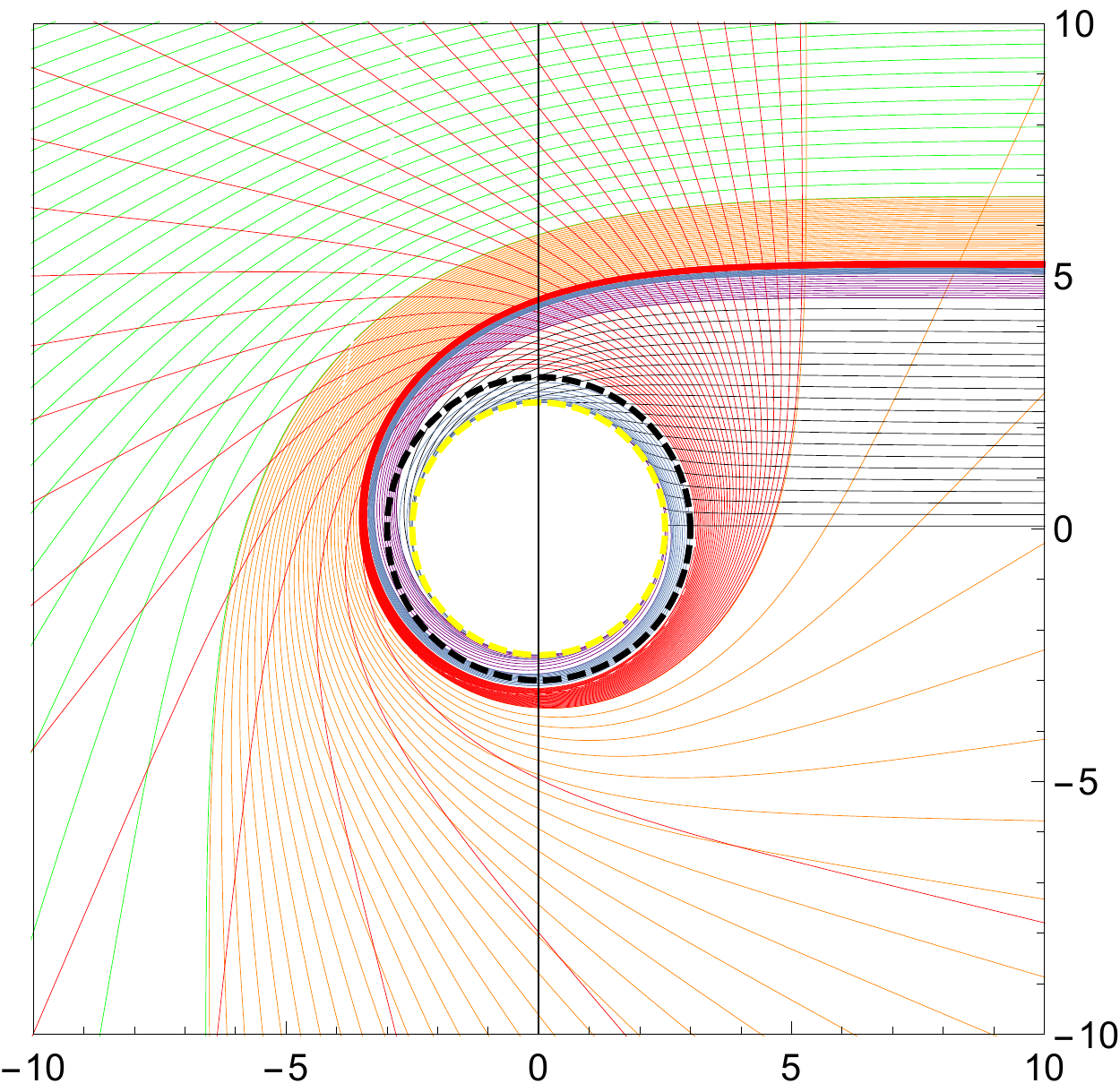}
\caption{Ray-tracing for BB solutions with $a=0$ (Schwarzschild case, left figure), $a=3/2$ (BB BH case, middle figure) and $a=5/2$ (BB traversable WH case, right figure), written all of them in terms of the radial function $r$. The region inside the event horizon is represented by a black circle, while the photon sphere radius is represented by the dashed thick black circumference. The observer is located on the far right of the screen, and the different regions of the image are associated to the number of crossings with the equatorial plane outside the critical curve: direct emission (green), lensed emission (orange), photon ring emission (red). In addition the retro-trajectories (trajectories emerging out of the photon sphere) are also depicted: retro-photon ring (blue), retro-lensed (purple) and retro-direct (black). The dashed yellow circumference denotes the radius of the (spherical) mouth in the WH case. The vertical black solid line represents the location of the accretion disk in all cases.}
\label{fig:raytra}
\end{figure*}

In Table \ref{table:I} we have displayed several impact parameters covering the ranges between the critical cases $a=0$ (Schwarzschild), $a=2$ (transition BH/WH) and $a=3$ (disappearance of the photon sphere), including two representative cases of the BH/WH solutions, $a=3/2$, $a=5/2$, respectively, to be later used in the illustration of the ray-tracing images as well as for the optical appearance of the BB solutions when illuminated by accretion disks.  There are several aspects to be underlined in the modifications of the light rays' impact parameters as compared to the Schwarzschild solution. First, for $b>b_c$ one can note a broadening in the range of impact parameters contributing to the direct/lensed/photon ring emissions as we increase the BB parameter $a$. This therefore leads to wider lensing/photon rings, and supposedly would contribute to enhance the corresponding luminosity in these regions. Second, in the BH case the impact parameter region of the retro-orbits ($b<b_c$) narrows for the retro-direct emission, but broadens for both the retro-lensed and retro-photon ring until the WH branch is reached ($a=2$), where the tendency is reversed until the limit $a=3$ is attained.

Focusing for instance on the photon ring ones, one sees a sharp increase in the width of the impact parameter region when moving from the BH to the WH configurations ($a=2$), as allowed by the uncloaking of the wormhole mouth at a radius $r_{th}=a>2M$. As the BB parameter $a$ is further increased the contribution of the retro-direct emission increases, while those of the retro-lensed and retro-photon ring decreases, but the total impact parameter region (i.e. joint emission from both kinds of contributions) is slightly increased for all the trajectories.

To illustrate this general discussion, let us take two representative samples of the BH/WH configurations, namely, $a=3/2$ and $a=5/2$, and integrate the geodesic equation (\ref{eq:geoBB}) for a bunch of light rays spanning the relevant region of impact parameters for these three plus three kinds of orbits. The corresponding results are depicted in Fig.\ref{fig:raytra} for values of the impact parameter $b \in (0,10)$, alongside its comparison with the known results of the Schwarzschild case ($a=0$), and  we point out that the observer's screen is located at the far right side of this plot in all these cases. In these figures one can clearly see the direct (green), lensed (orange) and photon ring (red) trajectories outside the photon sphere ($r=3M$, dashed black), and the impact parameter's range they correspond to. In the BH case with $a=3/2$ (middle figure) the enhance in the impact parameter's range as compared to the Schwarzschild solution regarding these trajectories is barely visible, though it is there, being much more noticeable in the WH case (right figure). In addition, we have plotted the retro-photon ring (blue), retro-lensed (yellow), and retro-direct (black), originated from inside the photon sphere, $b<b_c$. In the BH case, $a=3/2$, these contributions would intersect the event horizon at $r=2M$ (black circle), while in the WH case, $a=5/2$, such trajectories corresponds to those originated from the throat (purple dashed circumference). All these retro-trajectories reach the observer's screen after circling the BB solution and exiting the photon sphere.

\begin{figure*}[t!]
\includegraphics[width=5.8cm,height=4.0cm]{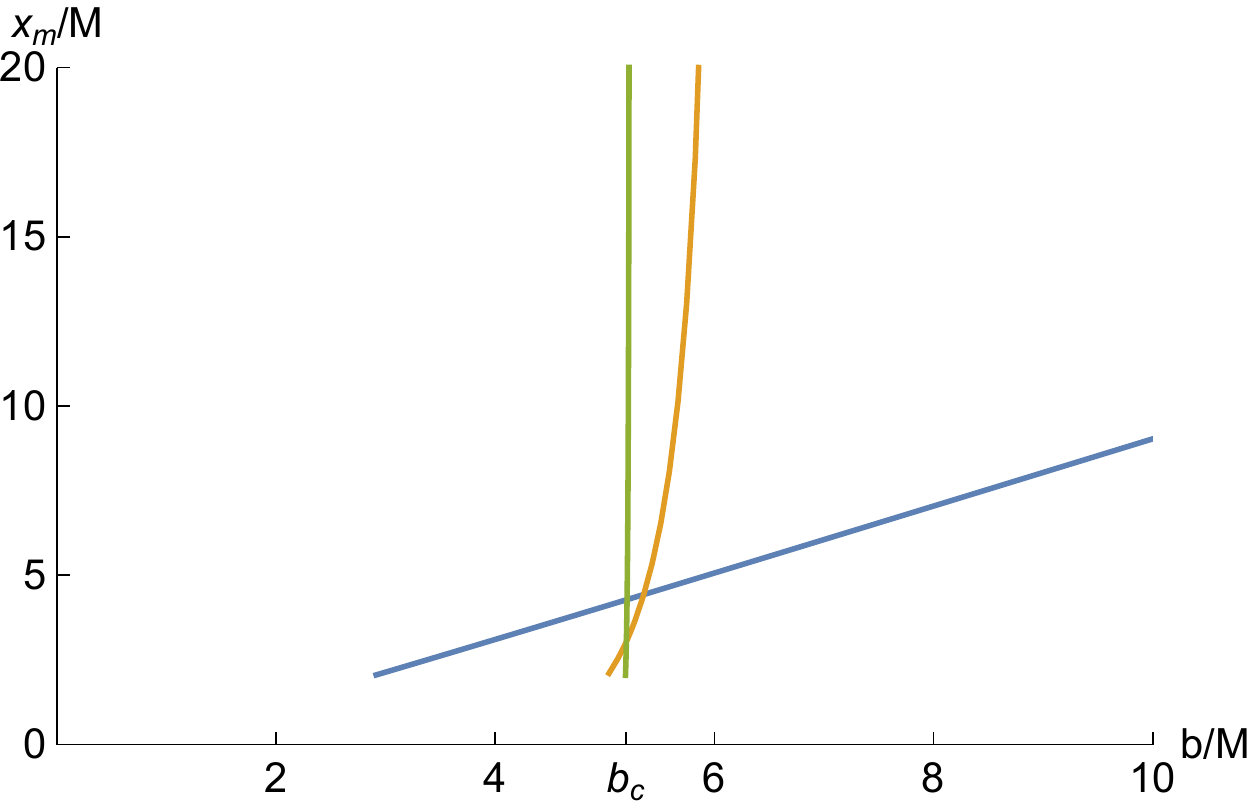}
\includegraphics[width=5.8cm,height=4.0cm]{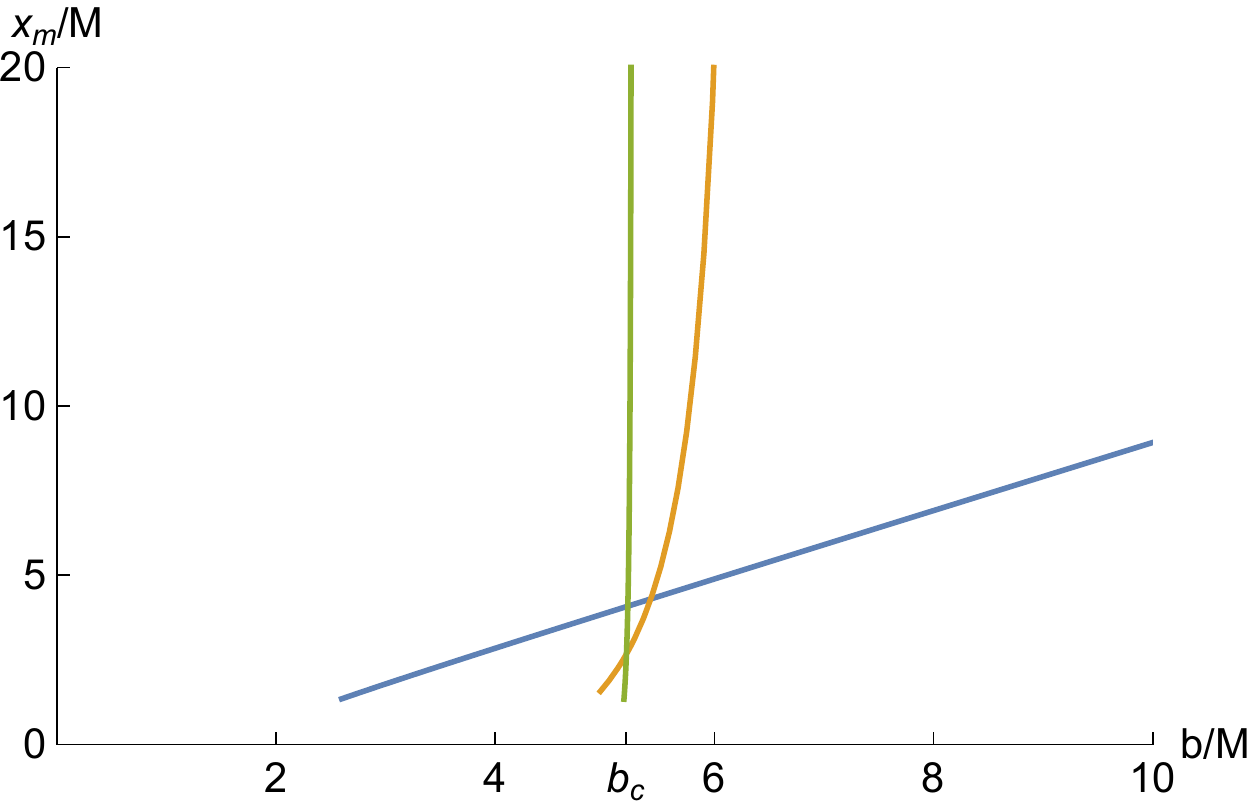}
\includegraphics[width=5.8cm,height=4.0cm]{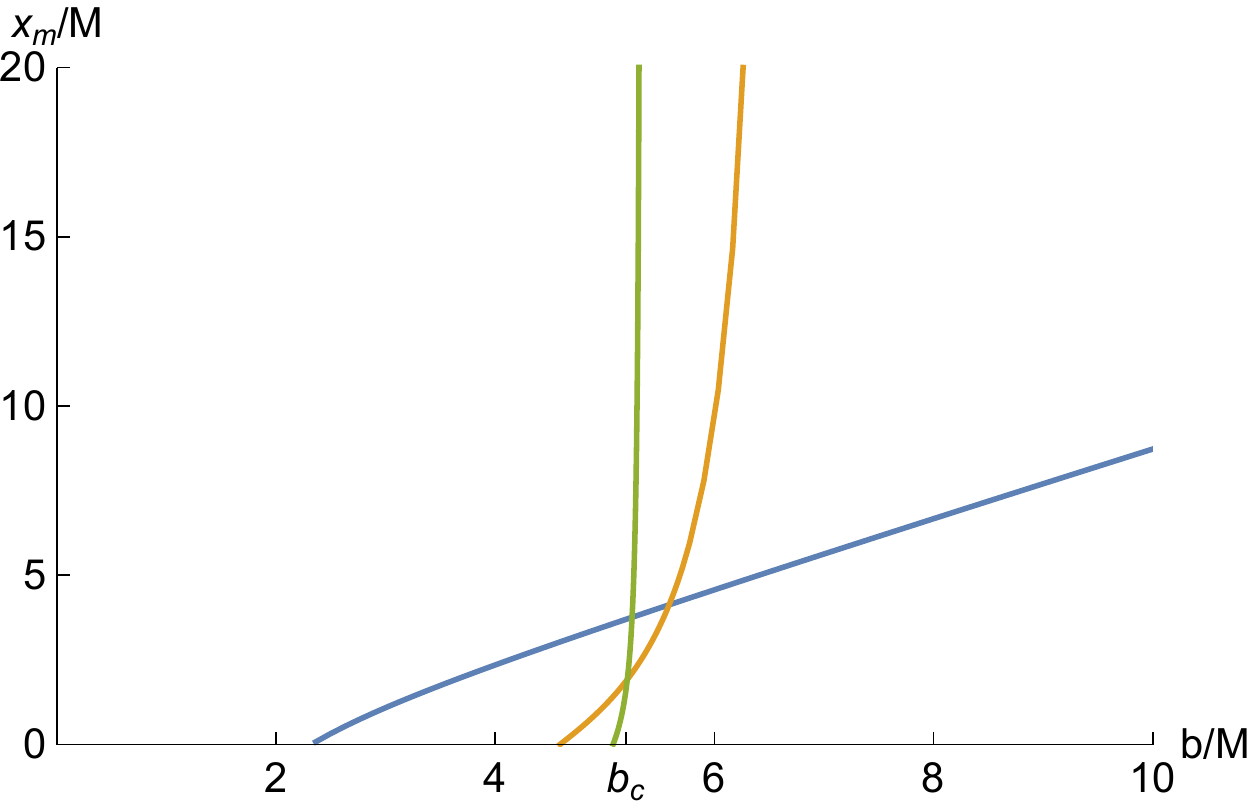}
\caption{The first three transfer functions $x_m(b)$ for a Schwarzschild solution ($a=0$, left), a BB solution of BH type ($a=3/2$, middle) and a BB solution of WH type ($a=5/2$, right). The colours depict direct (blue), lensed (orange) and photon ring (green) emissions, and include the contributions from the retro-orbits. The transfer functions get their minimum values in impact parameter at the point in which no intersections with the accretion disk are made ($n<1/4$), which are found at the horizon radius $x_h=\sqrt{4-a^2}$ for the Schwarzschild ($x_h=2, b\approx 2.852 $) and BB ($x_h \approx 1.322,b\approx 2.572$) solutions, while in the BB WH case it corresponds to the location of the wormhole throat ($x=0,b\approx 2.319$).}
\label{fig:transfer}
\end{figure*}

\section{Shadows from geometrically and optically thin accretion disks} \label{sec:IV}

We consider an optically and geometrically thin disk\footnote{This model describes accretion disks when the accretion rate is sub-Eddington with a very high opacity \cite{Page:1974he}, but leaves aside those with high mass accretion rates, for instance around supermassive black holes, which may effectively turn the accretion disk into an optically thin but geometrically thick one \cite{Riaz:2019bkv}.} surrounding our BB solution on the equatorial plane (so that the observer is located in the north pole), viewed face-on, and providing the main contribution to the observed intensity. To this end, we shall consider three toy models of accretion disks (therefore disregarding the complex  modelling of the plasma in realistic astrophysical scenarios, which requires full magneto-hydrodynamic simulations)  where the specific luminosities only depend on the radial coordinate $x$, and where the disks are assumed to emit isotropically, $I_{\nu}^{em}=I(x)$, with $\nu$ the emission frequency in the rest frame of the emission:

\begin{itemize}

\item Model I: the emission has a sharp peak at the innermost stable circular orbit (ISCO) for time-like observers, while vanishing in the region internal to it and falling off asymptotically to zero beyond of it. In the BB case, the ISCO radius reads $x_{isco}=\sqrt{36M^2-a^2}$ (or $r_{isco}=6M$). Therefore we model this emission profile as (taking $M=1$)
\begin{equation} \label{eq:ModI}
I_{em}^I=\left\lbrace\begin{array}{c}\frac{1}{(x-(x_{isco}-1))^2} \hspace{0.1cm} \text{if} \hspace{0.1cm} x \geq x_{isco} \\ 0 \hspace{2.0cm} \text{if} \hspace{0.1cm} x < x_{isco} \end{array}\right.
\end{equation}

\item Model II: the emission has a sharp peak at the unstable circular orbit location (\ref{eq:ps}), having a qualitatively similar central and asymptotic behaviour as Model I. This is described by
\begin{equation} \label{eq:ModII}
I_{em}^{II}=\left\lbrace\begin{array}{c}\frac{1}{(x-(x_{pr}-1))^3} \hspace{0.1cm} \text{if} \hspace{0.1cm} x \geq x_{pr} \\ 0 \hspace{1.8cm} \text{if} \hspace{0.1cm} x < x_{pr} \end{array}\right.
\end{equation}

\item Model III: the emission starts at the event horizon and falls off more smoothly to zero than in the previous two cases, being modelled as (in the WH case we take instead the wormhole mouth, $x_{th}=0$)
\begin{equation} \label{eq:ModIII}
I_{em}^{III}=\left\lbrace\begin{array}{c}\frac{\pi/2-\arctan[x-5]}{\pi/2-\arctan[x_{hor}-5]} \hspace{0.1cm} \text{if} \hspace{0.1cm} x \geq x_{hor} \\ 0 \hspace{2.6cm} \text{if} \hspace{0.1cm} x < x_{hor} \end{array}\right.
\end{equation}

\end{itemize}

The observed intensity on the receiver's screen is given by the gravitationally red-shifted emitted density (disregarding effects associated to absorption and reflection of light). Given the fact that $I_{\nu}/\nu^3$ is conserved along a photon trajectory, one finds that in the line element (\ref{eq:sss}) the observed intensity at a frequency $\nu'$ scales with respect to the emitted one as $I_{\nu'}^{ob}=A^{3/2}(x)I(x)$. Therefore, the total observed intensity will be the integration over the whole range of received frequencies as $I^{ob}=\int I_{\nu'}d\nu'$ or, in other words, $I^{ob}=A^2(x)I(x)$. In our ray-tracing setup developed in Sec. \ref{sec:III}, whenever any light ray backtracked from the observer's screen crosses the accretion disk plane it will pick up additional brightness from it depending on the number of orbits. Therefore, the total received luminosity will be the sum of all the intensities from all these crossings with the accretion disk as
\begin{equation}
I^{ob}(b)=\sum_m A^2I_{\vert_{x=x_m(b)}} \ ,
\end{equation}
where the so-called ``transfer function" $x_m(b)$ contains the information about the radius of the disk where a given light ray with impact parameter $b$ will have its  $m^{th}$-intersection with the disk (in the coordinate $x$). Moreover, its slope $dx/db$ defines the (de)magnification of the image for the different types of emission (direct/lensed/photon ring). As it can be seen in Fig. \ref{fig:transfer}, the transfer function for the direct emission ($m=1$) has a constant nearly unit slope, while the lensed ($m=2$) and photon ring ($m=3$) emissions have quite a large slope, meaning that they are highly demagnified. Further crossings with the disk will lead to exponentially demagnified images \cite{Gralla:2019xty}, so they can be safely ignored .

\begin{table}[t!]
\begin{tabular}{|c|c|c|c|c|}
\hline
Model/$I^{ob}$   & BB parameter  & $I_{Direct}$ & $I_{Lensed}$ &  $I_{Photon}$  \\ \hline
\multirow{3}{*}{Model I}   & $a=0$  & $0.949$   & $0.0494$   & $0.00182$  \\ \cline{2-5}
                           & $a=3/2$  & $0.942$   & $0.0549$   & $0.00264$  \\ \cline{2-5}
                           & $a=5/2$ & $0.930$  & $0.0618$   & $0.00839$  \\ \hline
\multirow{3}{*}{Model II}  & $a=0$  & $0.901$   & $0.0940$   & $0.00412$  \\ \cline{2-5}
                           & $a=3/2$ & $0.878$   & $0.1145$   & $0.00731$  \\ \cline{2-5}
                           & $a=5/2$ & $0.798$   & $0.1760$   & $0.02589$  \\ \hline
\multirow{3}{*}{Model III} & $a=0$  & $0.926$   & $0.0715$   & $0.00281$  \\ \cline{2-5}
                           & $a=3/2$ & $0.908$   & $0.0871$   & $0.00506$  \\ \cline{2-5}
                           & $a=5/2$ & $0.877$   & $0.1094$  & $0.01398$  \\ \hline
\end{tabular}
\caption{Contributions of the direct, lensed, and photon ring emissions (including both $b \gtrless b_c$ trajectories) to the (normalized) total emission as seen by the observer, for the Schwarzschild ($a=0$), BB BH ($a=3/2$) and WH ($a=5/2$) solutions.}
\label{table:II}
\end{table}

\begin{figure*}[t!]
\centering
\includegraphics[width=5.9cm,height=5.0cm]{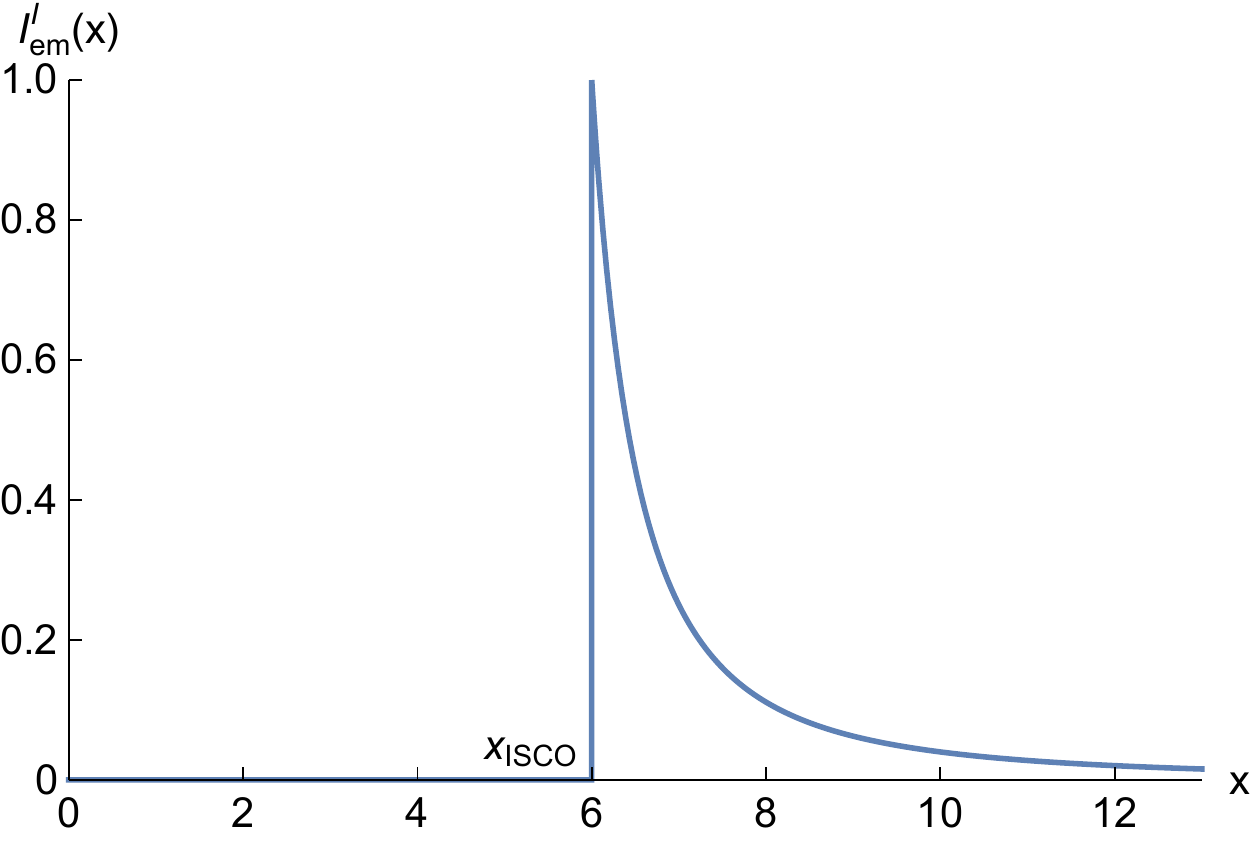}
\includegraphics[width=5.9cm,height=5.0cm]{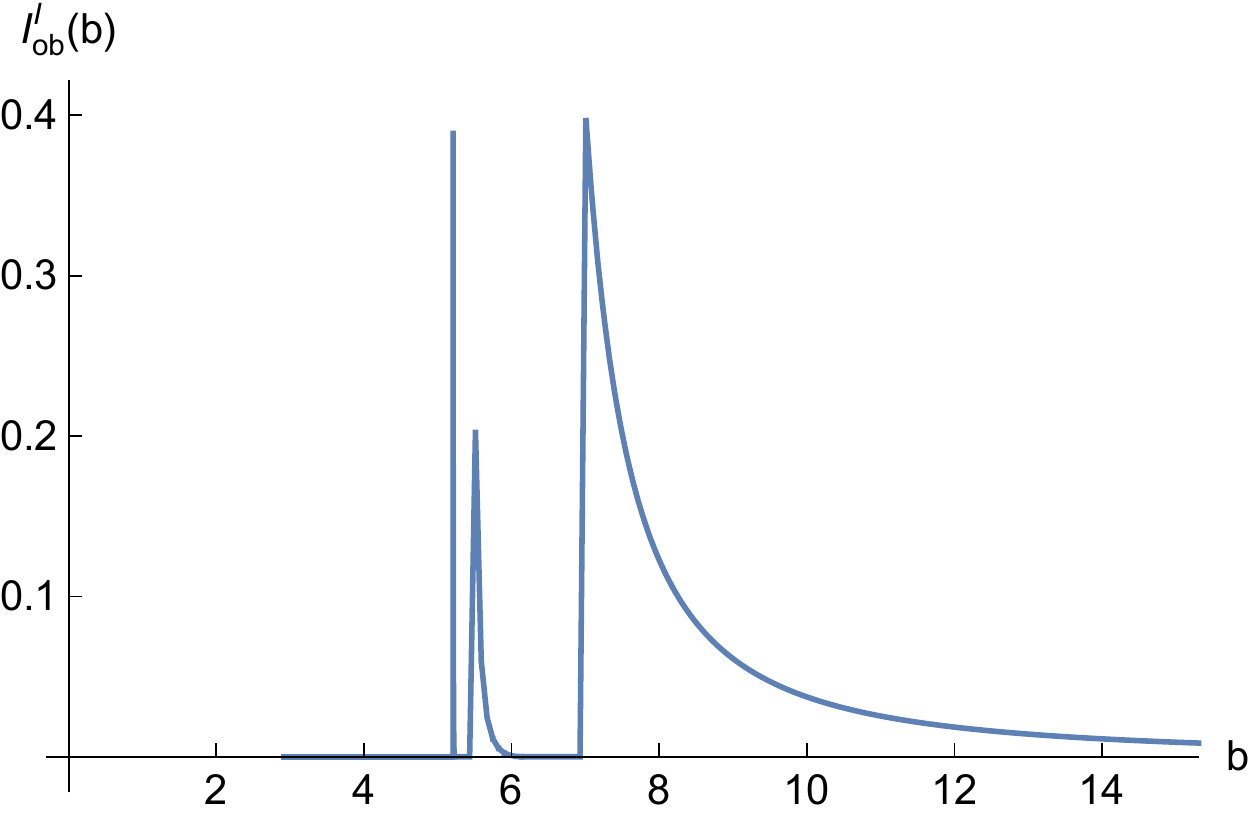}
\includegraphics[width=5.9cm,height=5.0cm]{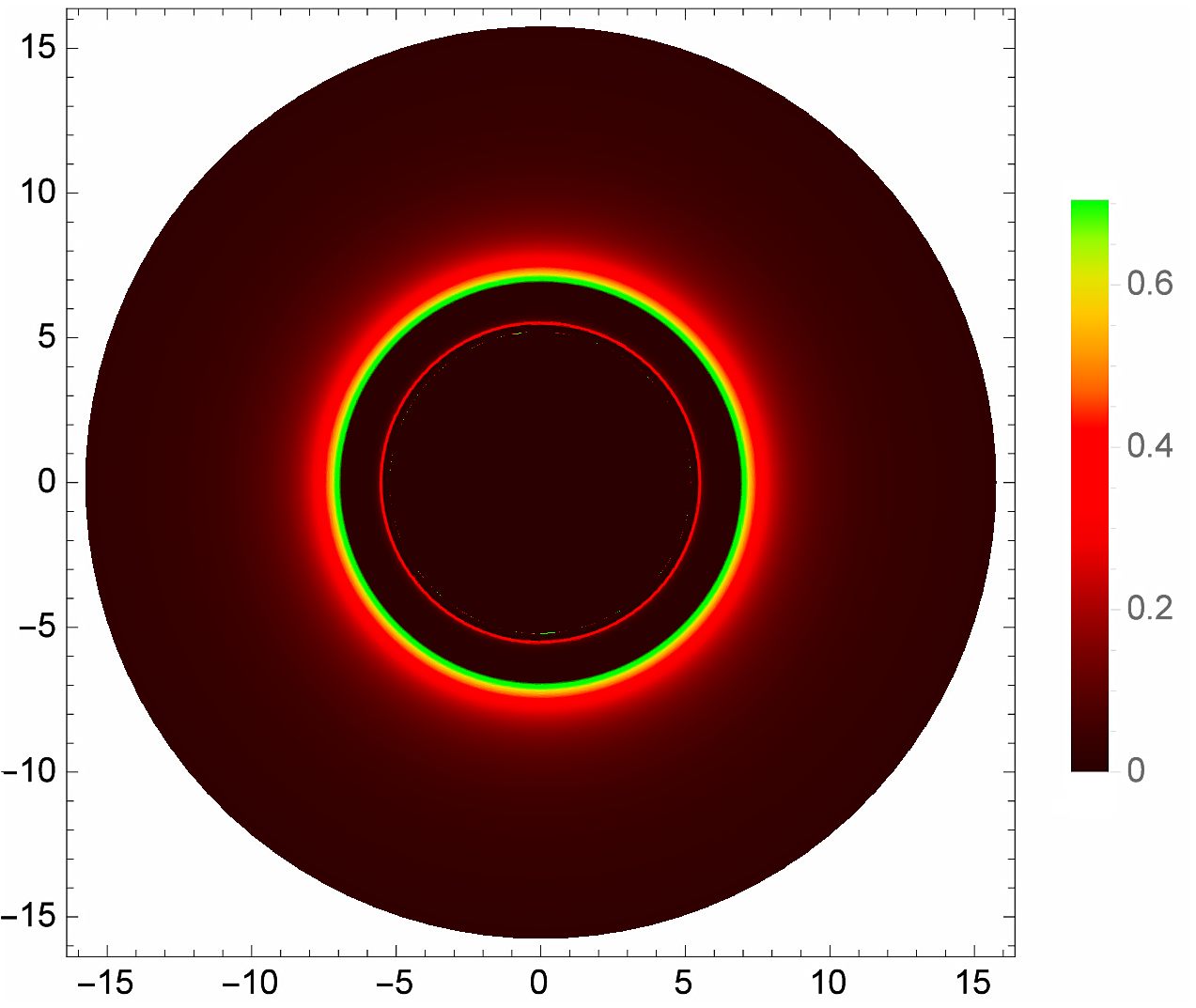}
\includegraphics[width=5.9cm,height=5.0cm]{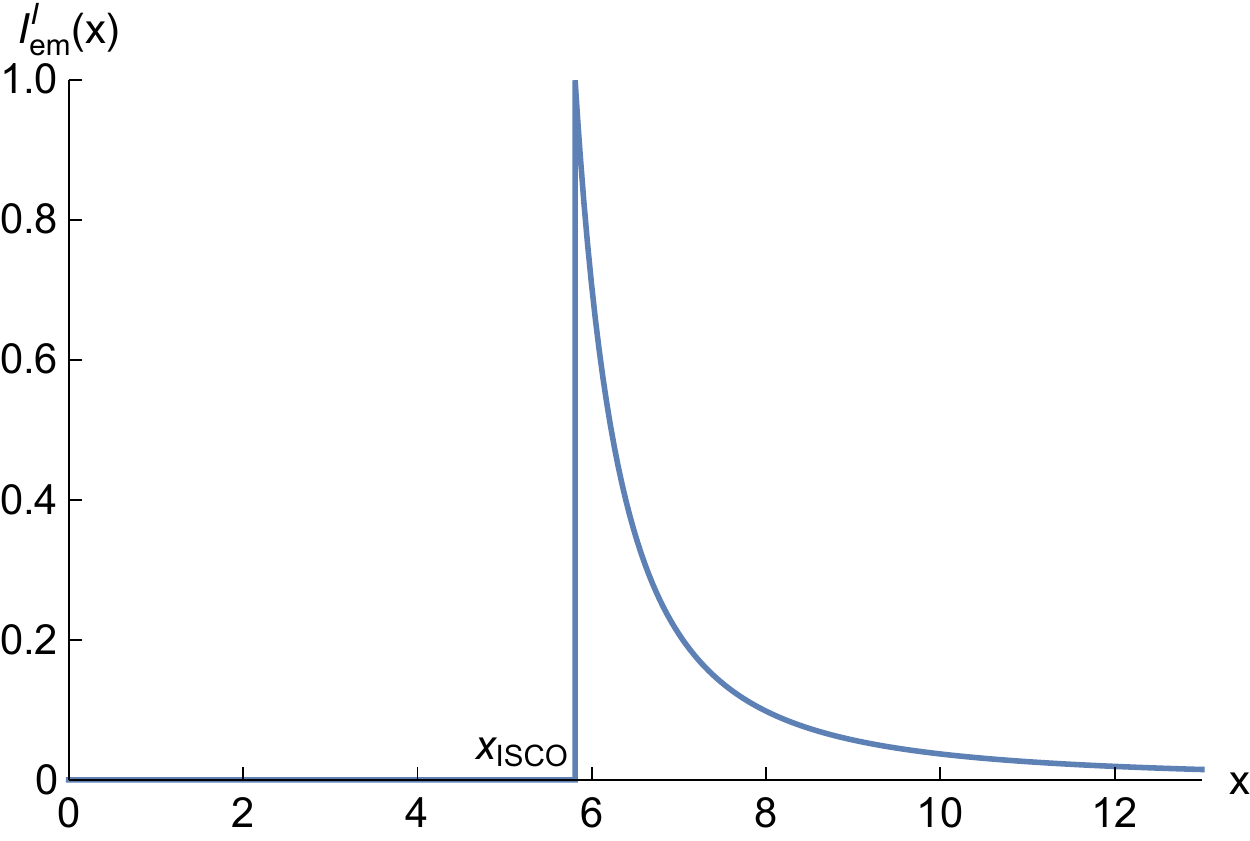}
\includegraphics[width=5.9cm,height=5.0cm]{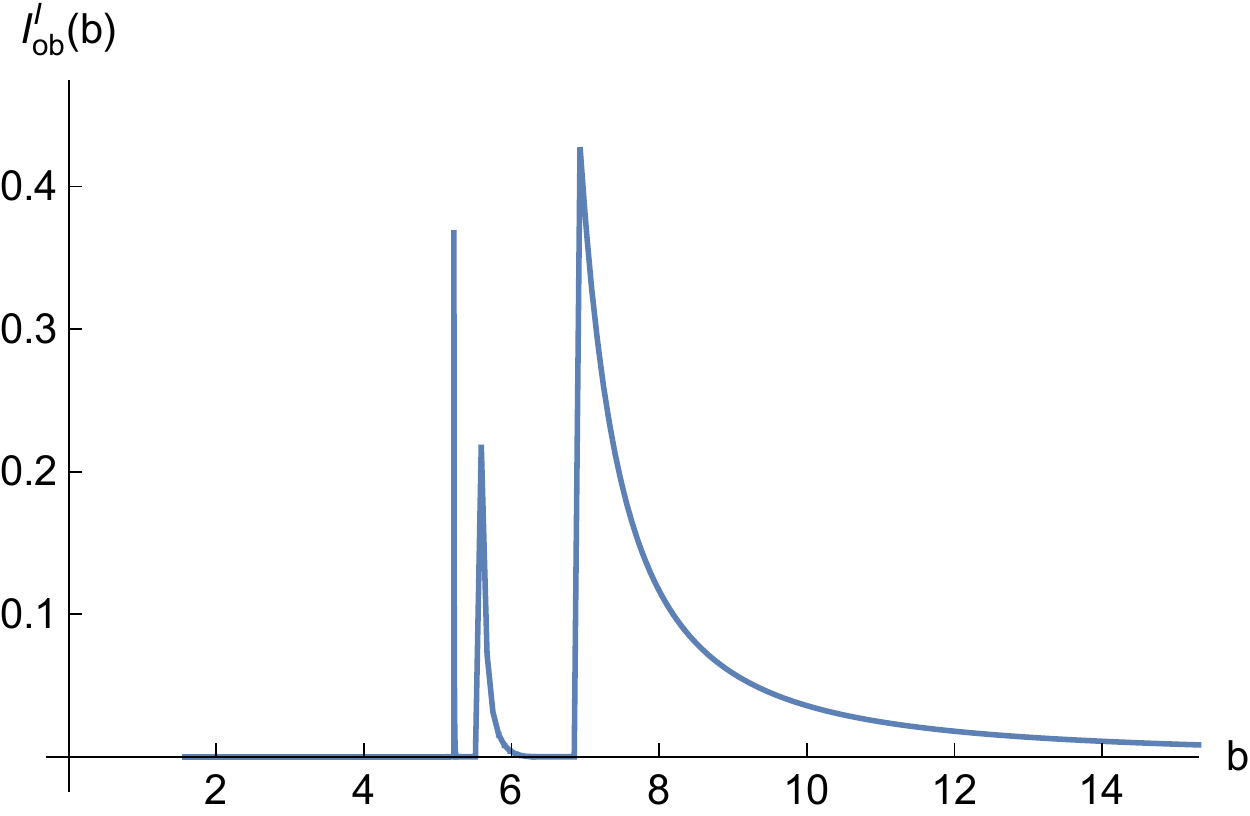}
\includegraphics[width=5.9cm,height=5.0cm]{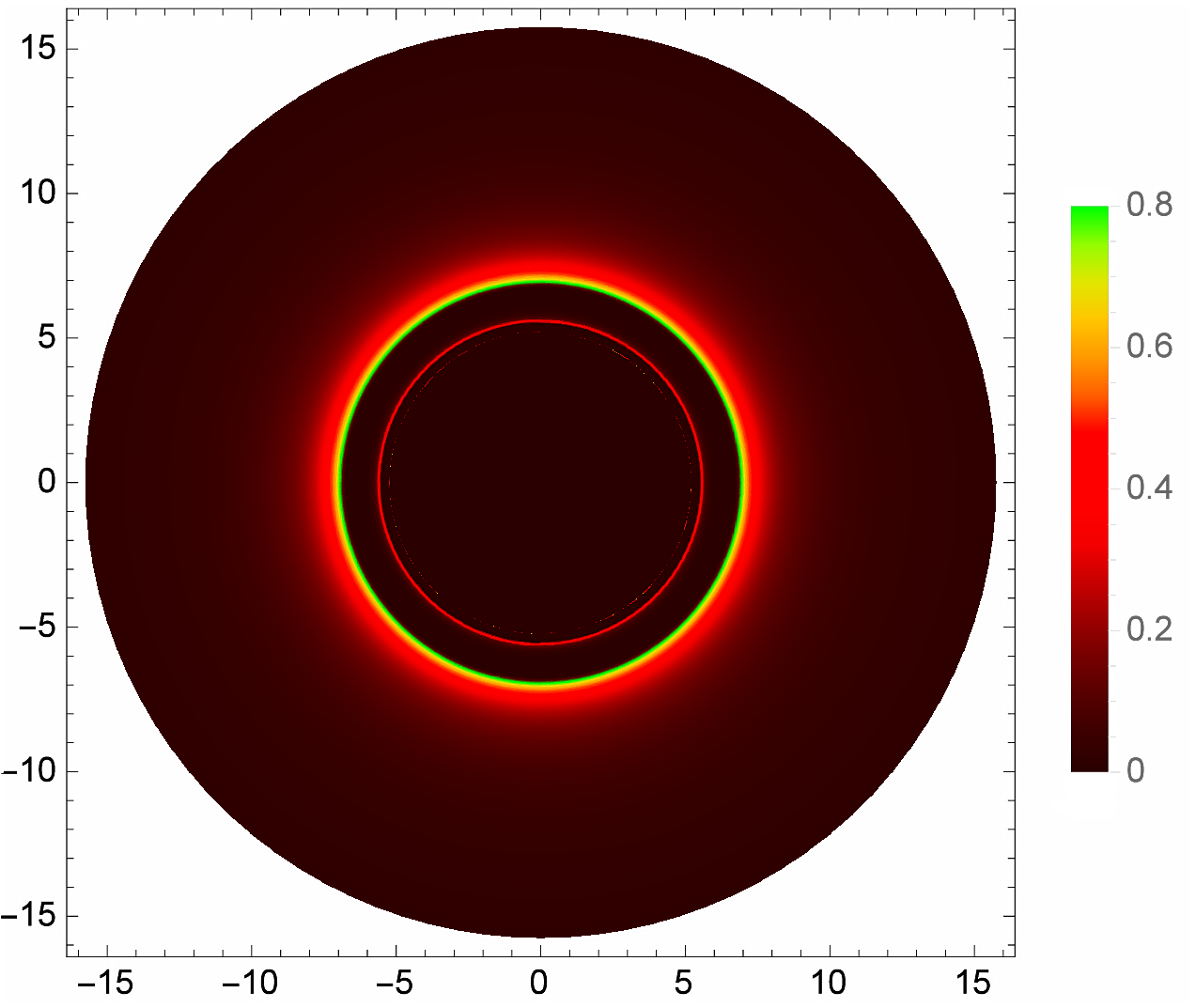}
\includegraphics[width=5.9cm,height=5.0cm]{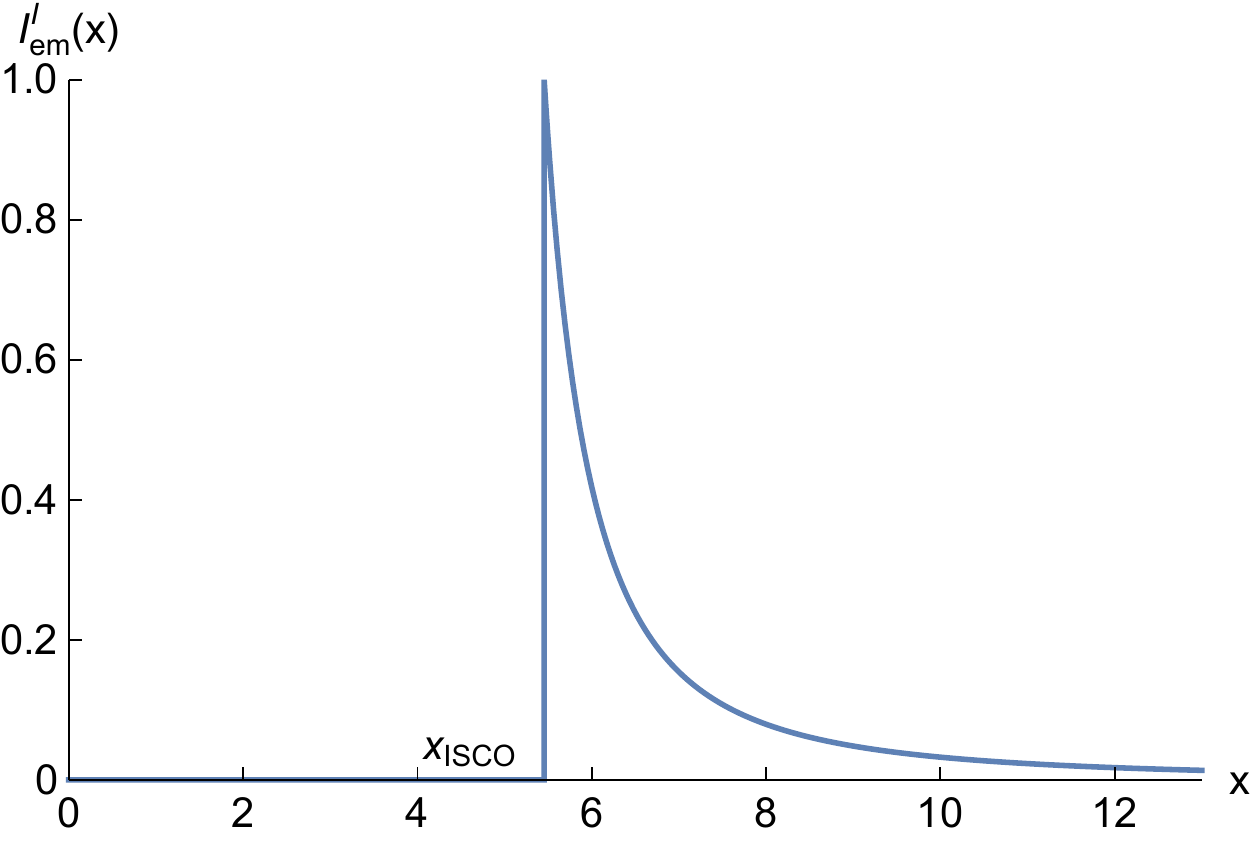}
\includegraphics[width=5.9cm,height=5.0cm]{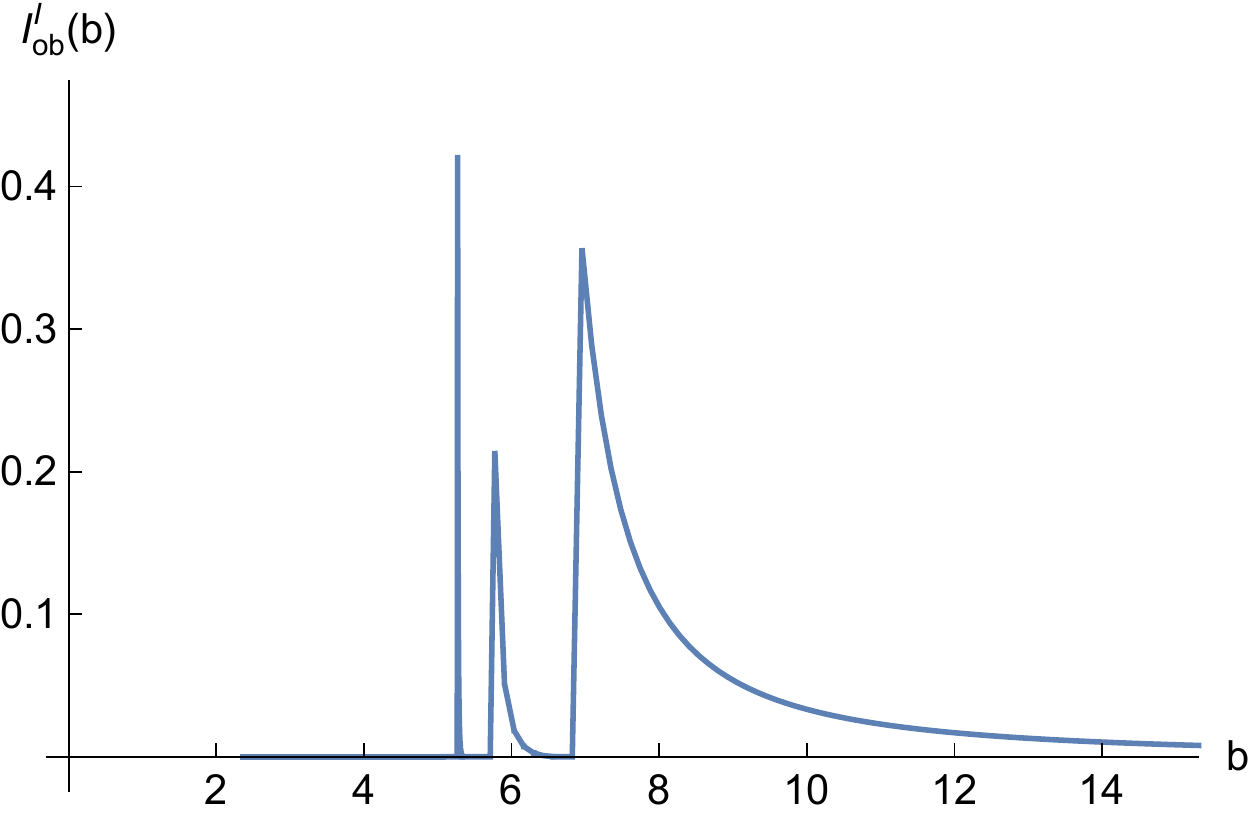}
\includegraphics[width=5.9cm,height=5.0cm]{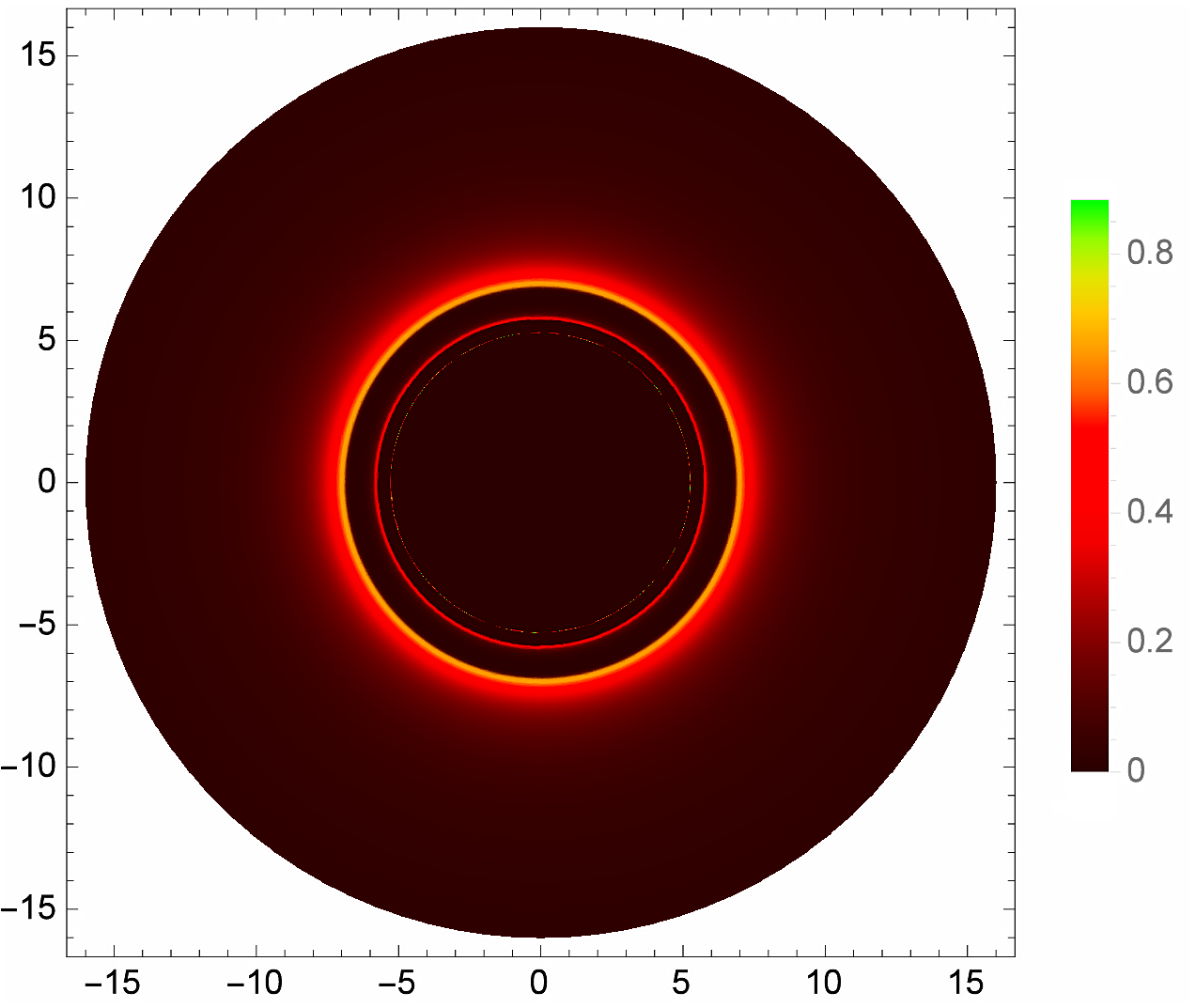}
\caption{The observational appearance of the BB solutions within accretion disk Model I (\ref{eq:ModI}) with $a=0$ (Schwarzschild, top), $a=3/2$ (BH case, middle) and $a=5/2$ (WH case, bottom), viewed from a face-on orientation. From left to right one finds the emitted profile, the observed one, and the optical appearance (in celestial coordinates) for a given BB solution. In the emission profiles we have made use of the radial coordinate $x$, related to the radial function as $r^2=x^2+a^2$ [recall Eq.(\ref{eq:BBBline})], which for the Schwarzschild case reads simply as $r^2 \approx x^2$. The observed profiles and the optical appearance are plotted as functions of the impact parameter. In these plots $x_{isco}=\sqrt{36M^2-a^2}$ is the radius of the innermost stable circular orbit for time-like observers, at which the emission of this Model I starts.}
\label{fig:model1}
\end{figure*}

\begin{figure*}[t!]
\centering
\includegraphics[width=5.9cm,height=5.0cm]{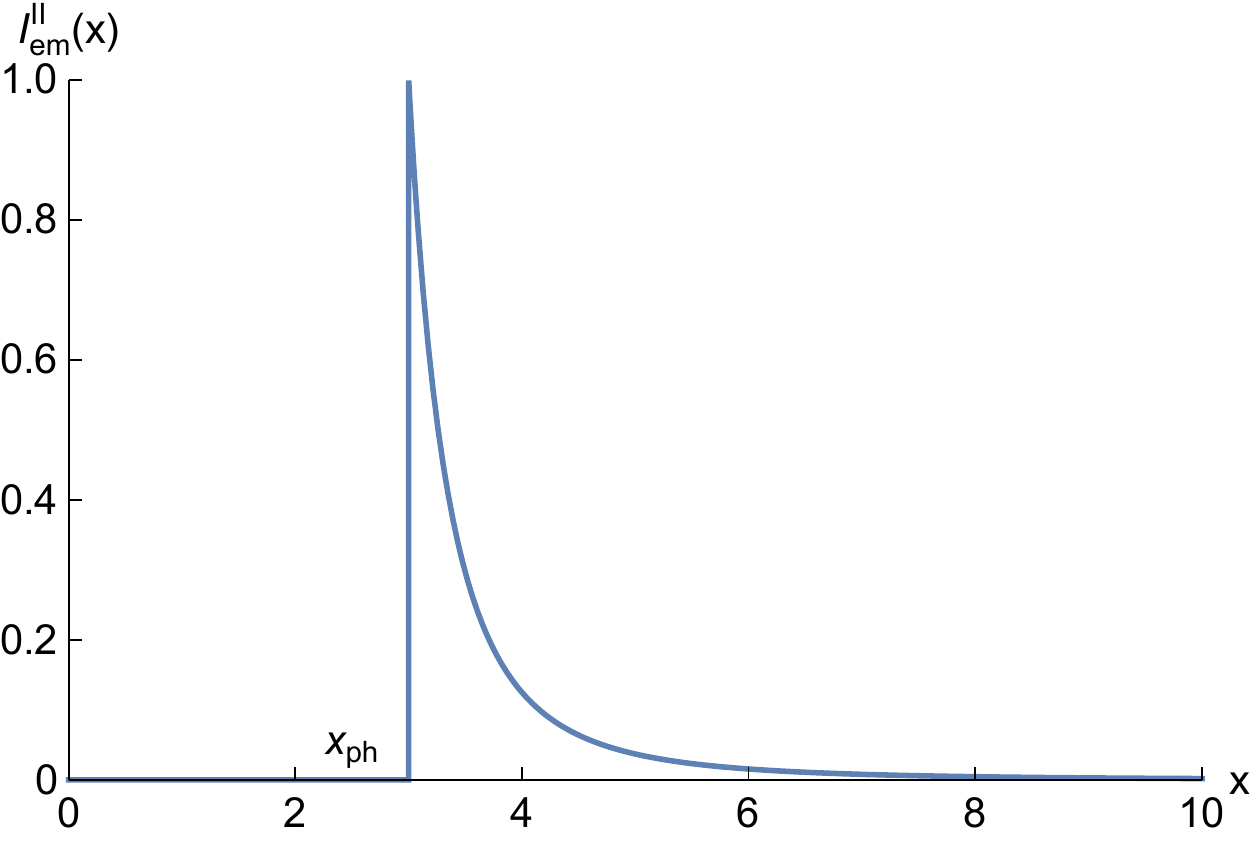}
\includegraphics[width=5.9cm,height=5.0cm]{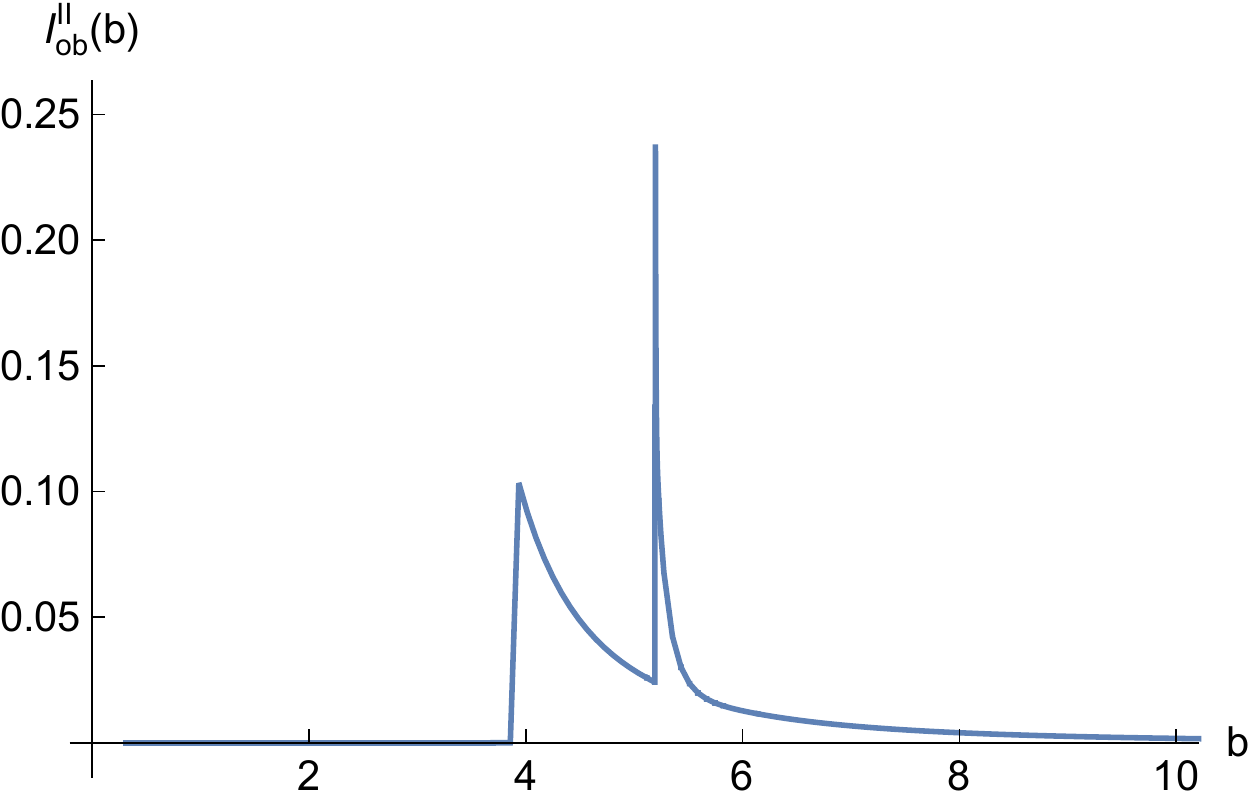}
\includegraphics[width=5.9cm,height=5.0cm]{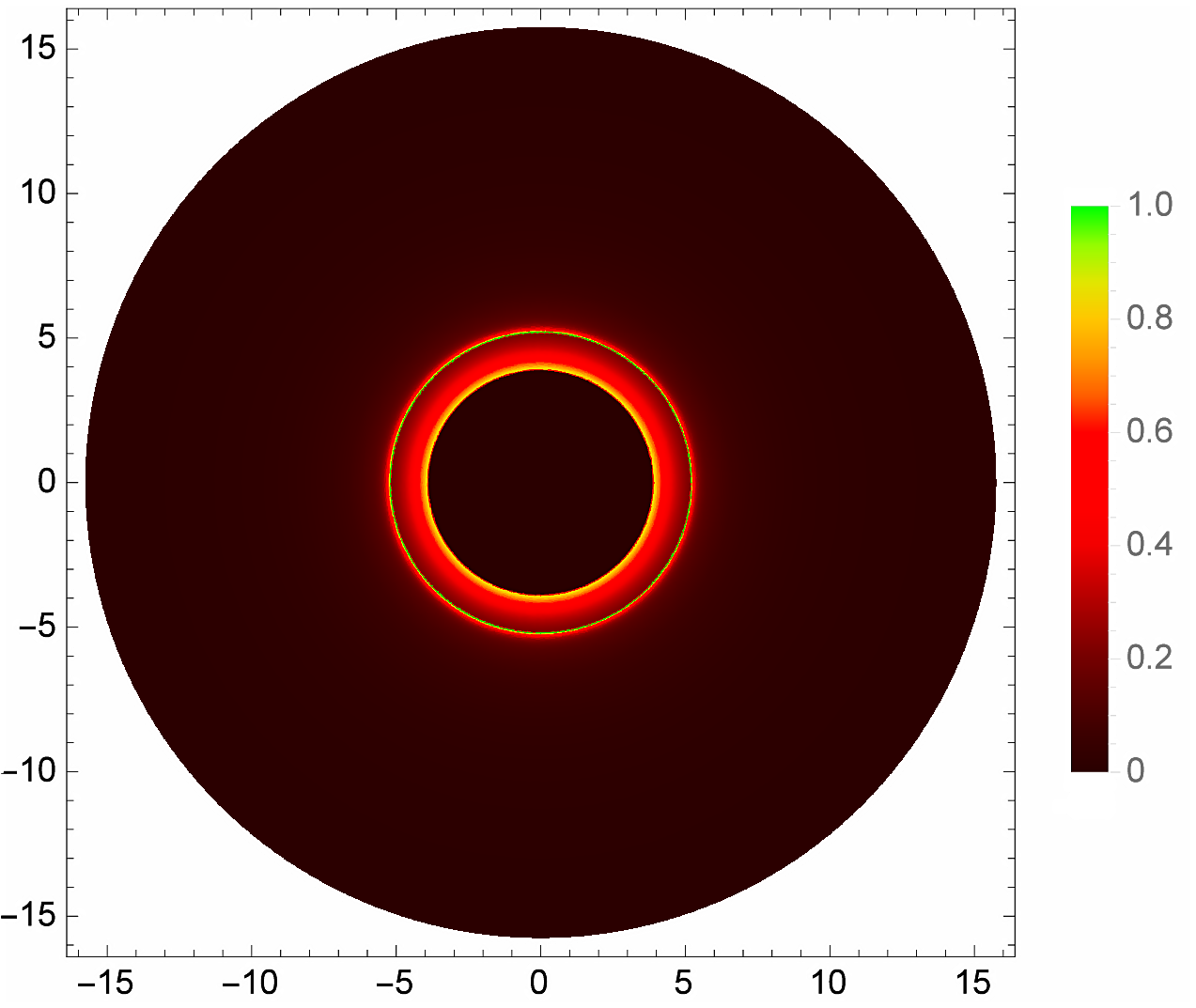}
\includegraphics[width=5.9cm,height=5.0cm]{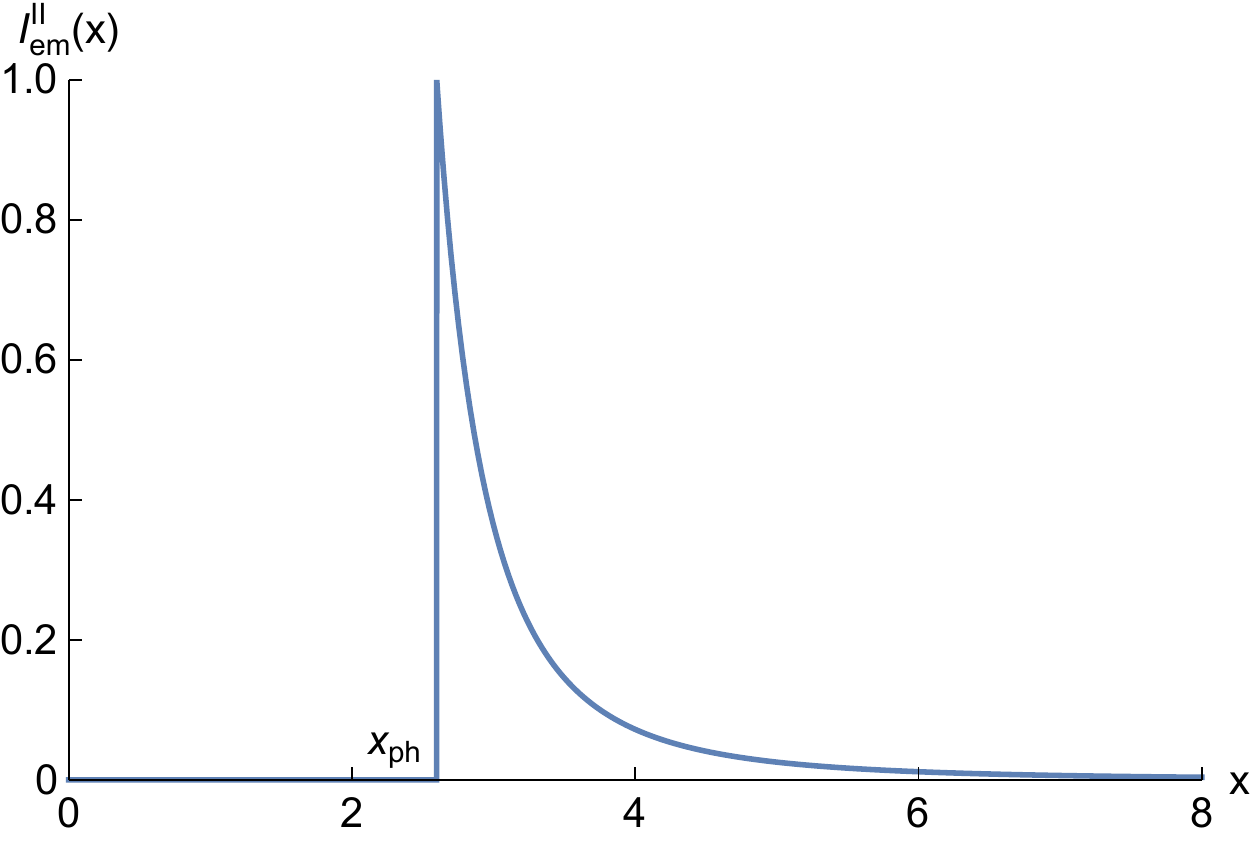}
\includegraphics[width=5.9cm,height=5.0cm]{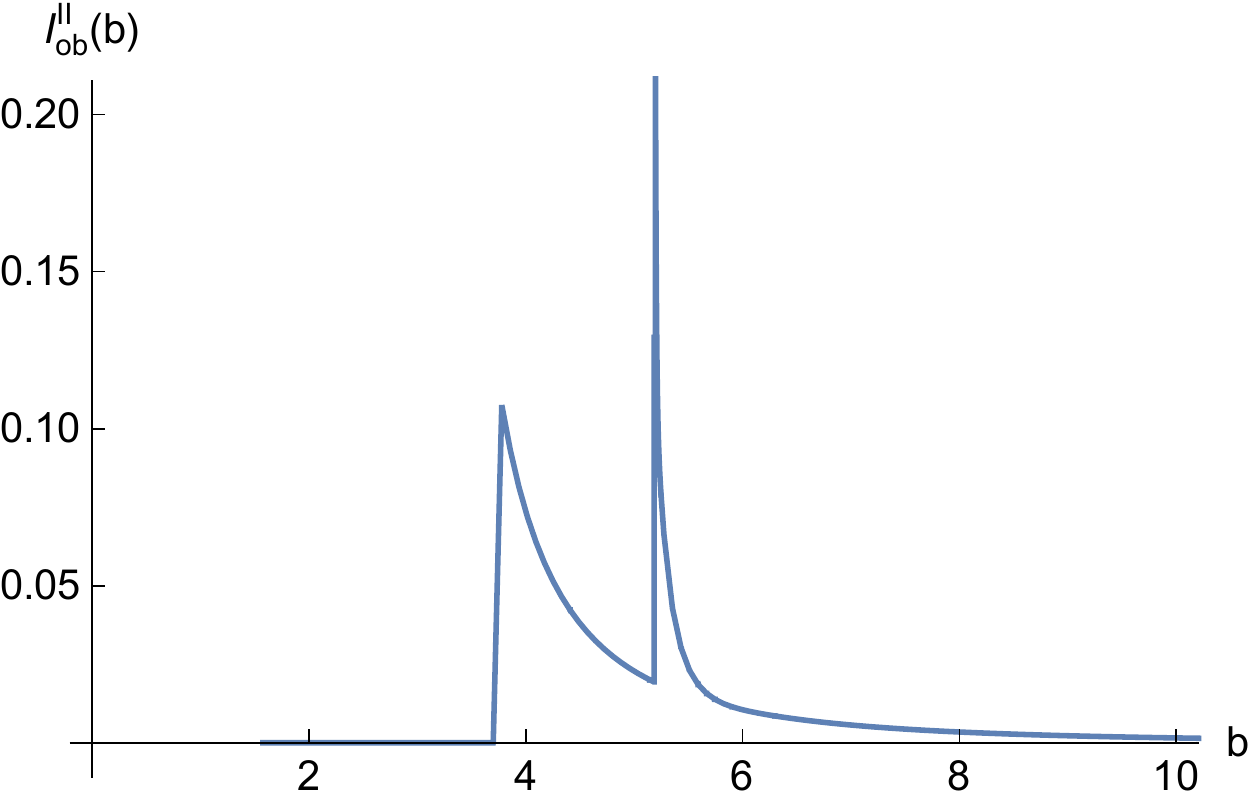}
\includegraphics[width=5.9cm,height=5.0cm]{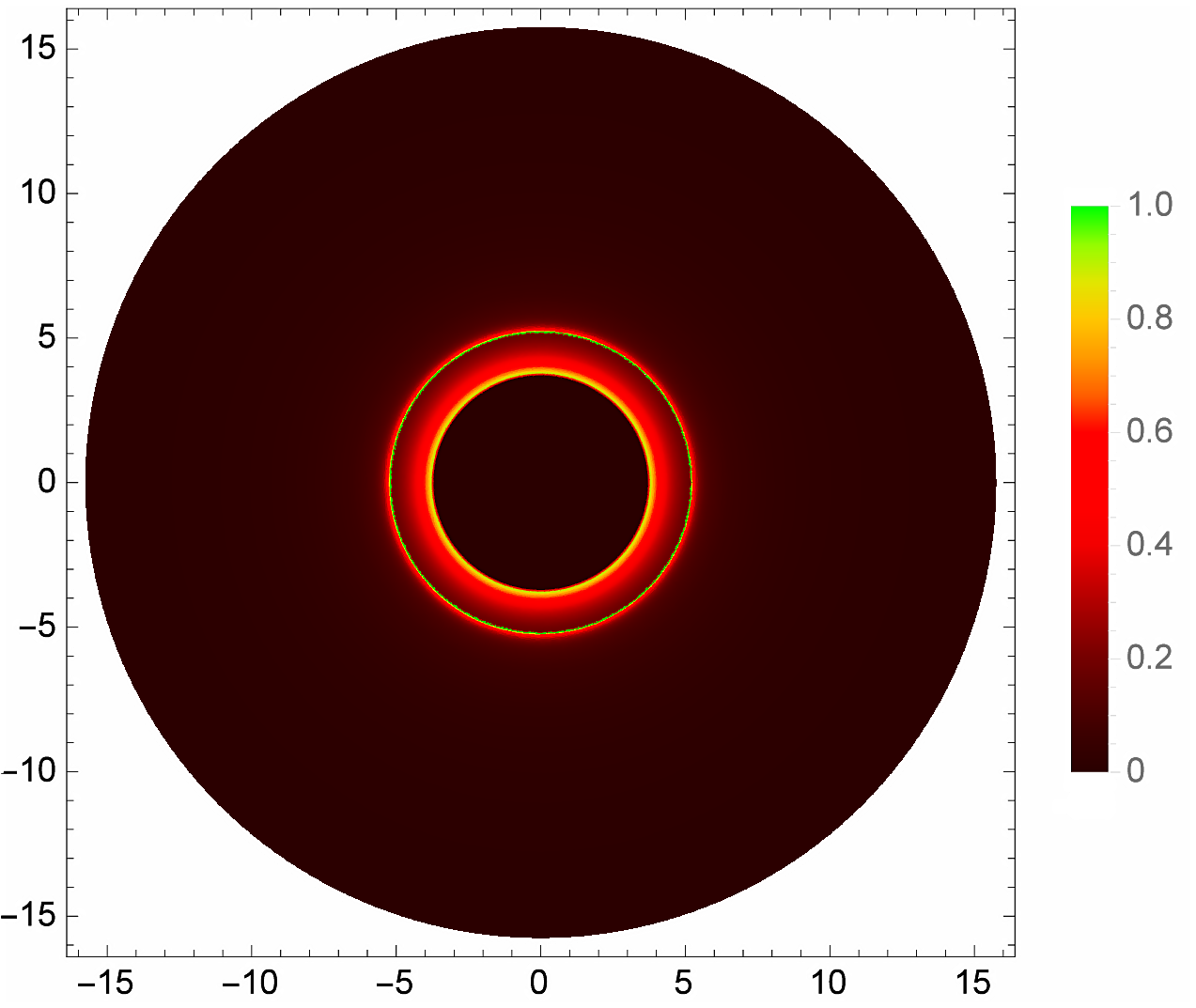}
\includegraphics[width=5.9cm,height=5.0cm]{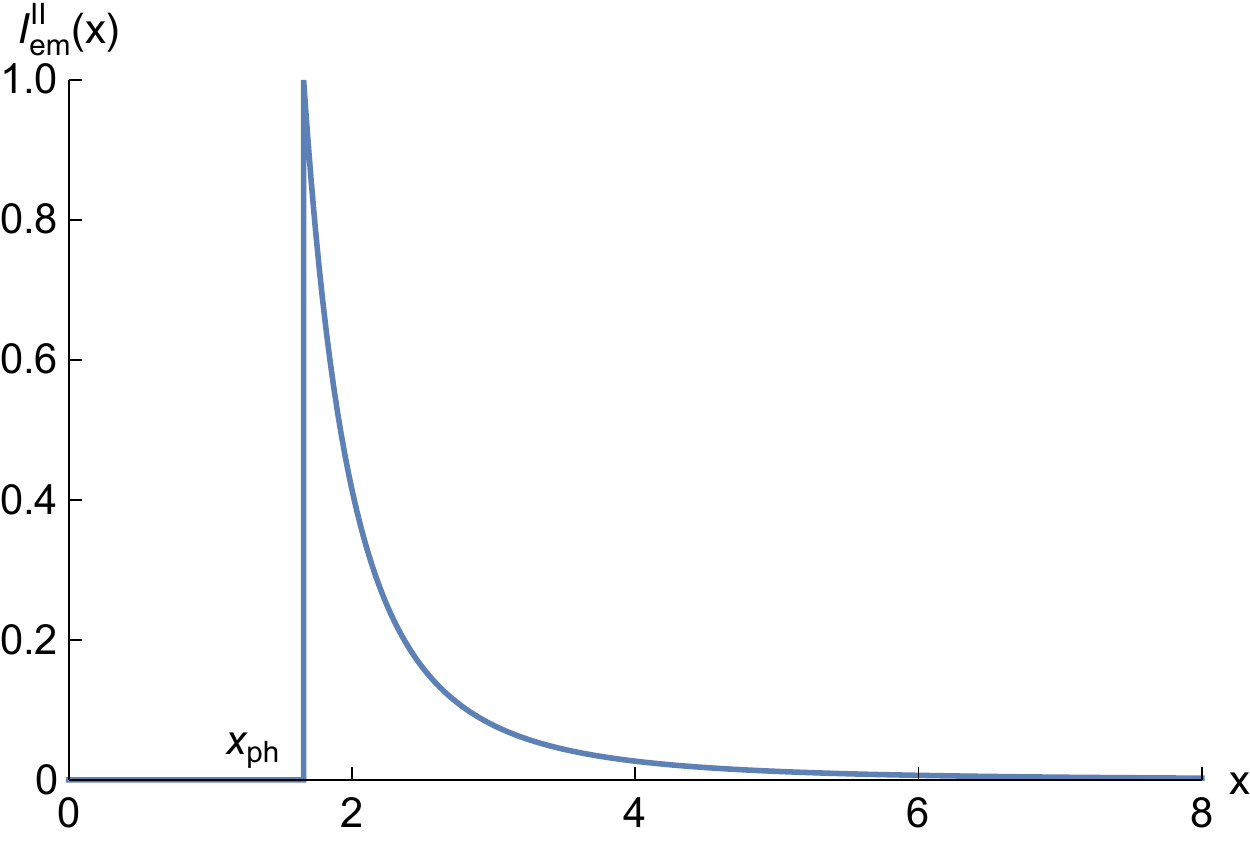}
\includegraphics[width=5.9cm,height=5.0cm]{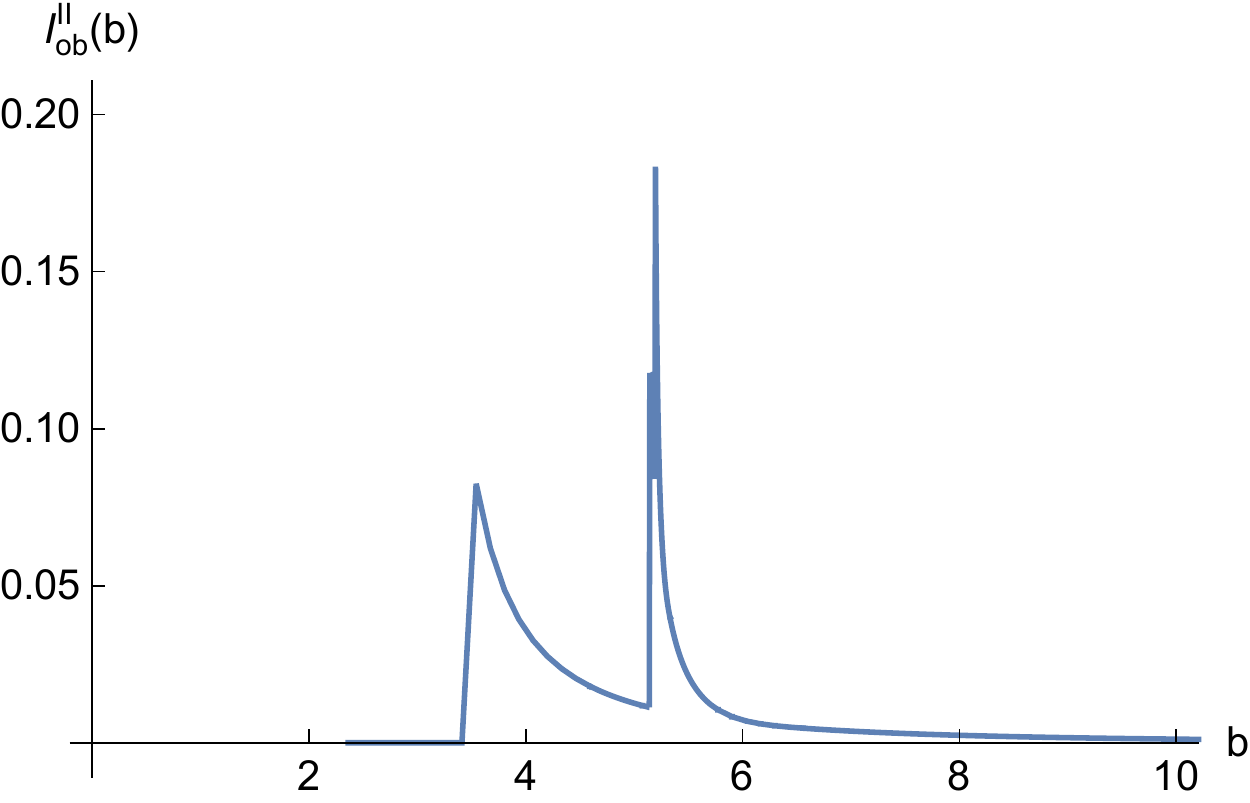}
\includegraphics[width=5.9cm,height=5.0cm]{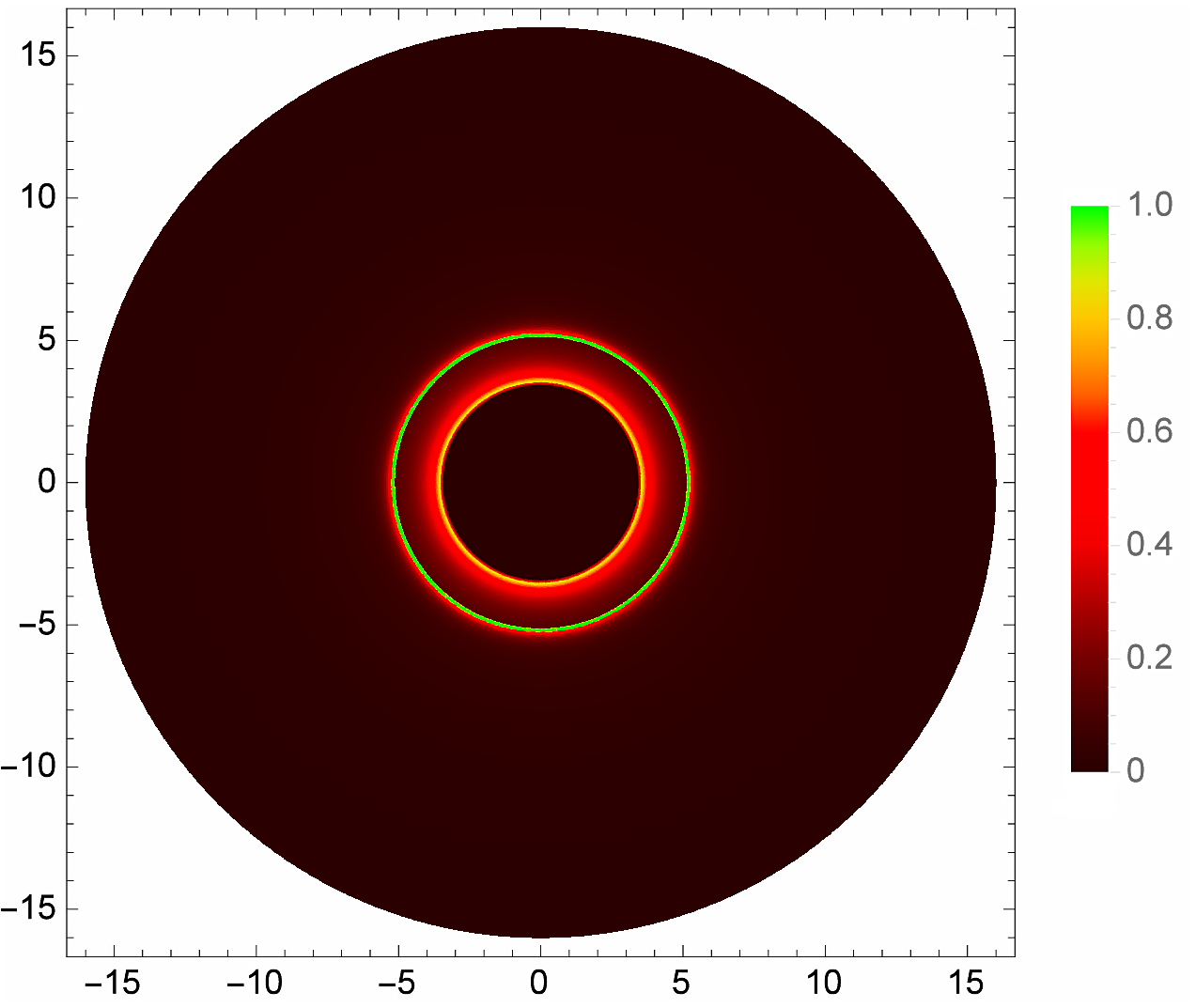}
\caption{The observational appearance of the BB solutions within accretion disk Model I (\ref{eq:ModI}) with $a=0$ (Schwarzschild, top), $a=3/2$ (BH case, middle) and $a=5/2$ (WH case, bottom), viewed from a face-on orientation, and with a similar notation as in Fig. \ref{fig:model1}. In these plots $x_{ph}=\sqrt{9M^2-a^2}$ is the photon sphere radius, at which the emission of this Model II starts.}
\label{fig:model2}
\end{figure*}

\begin{figure*}[t!]
\centering
\includegraphics[width=5.9cm,height=5.0cm]{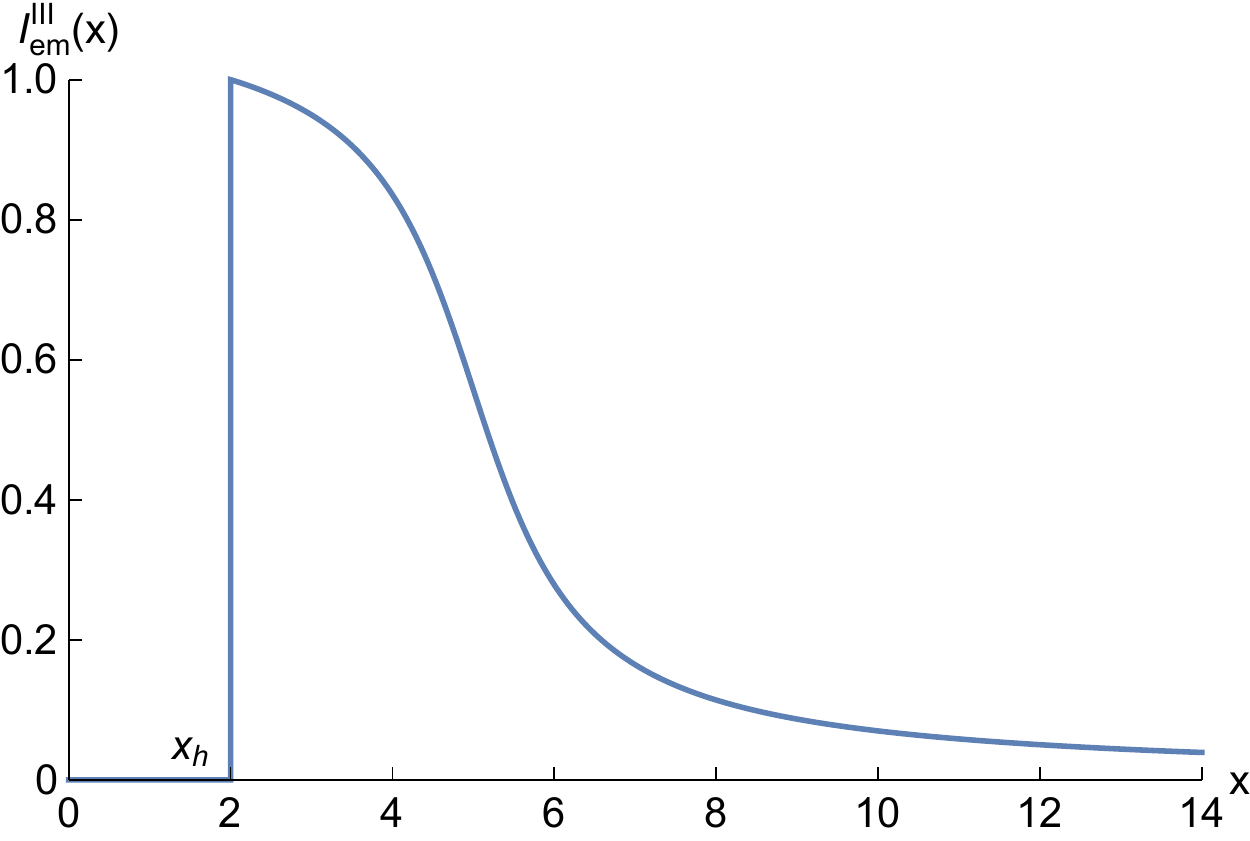}
\includegraphics[width=5.9cm,height=5.0cm]{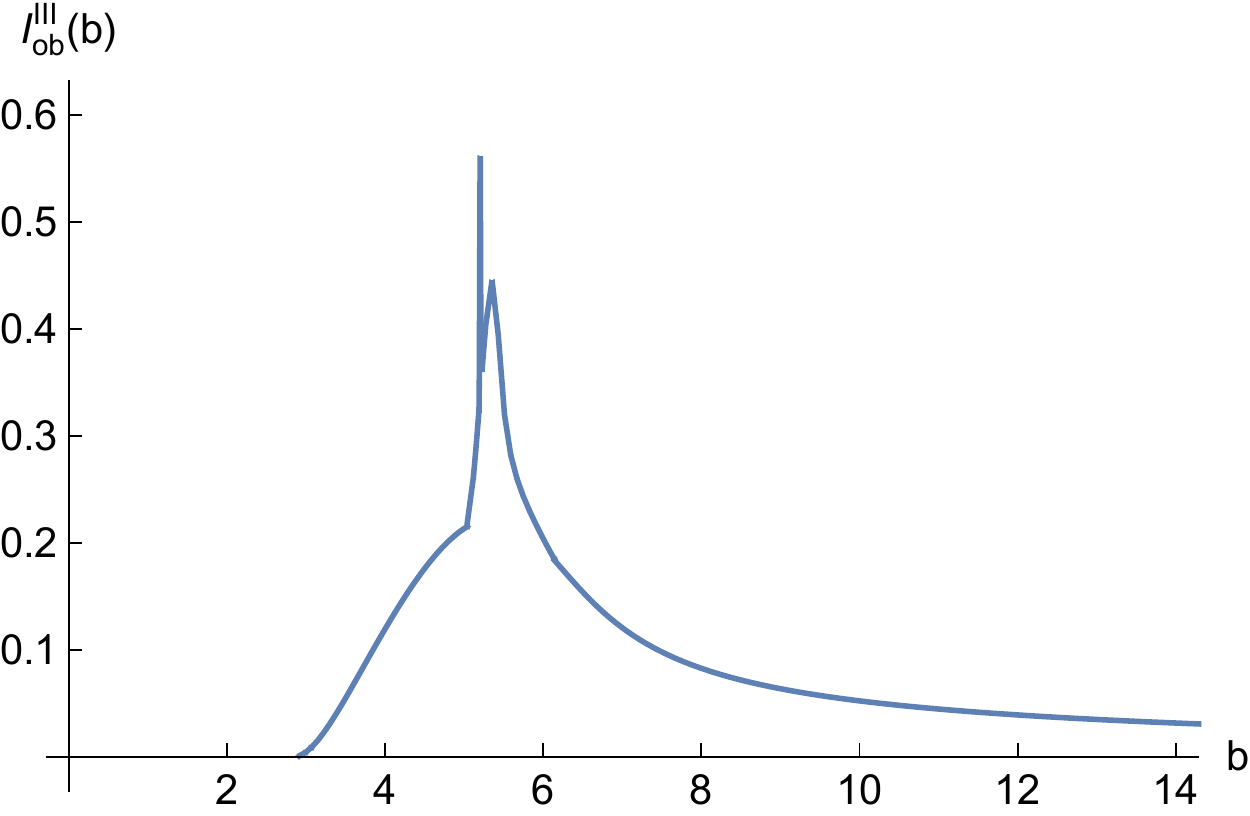}
\includegraphics[width=5.9cm,height=5.0cm]{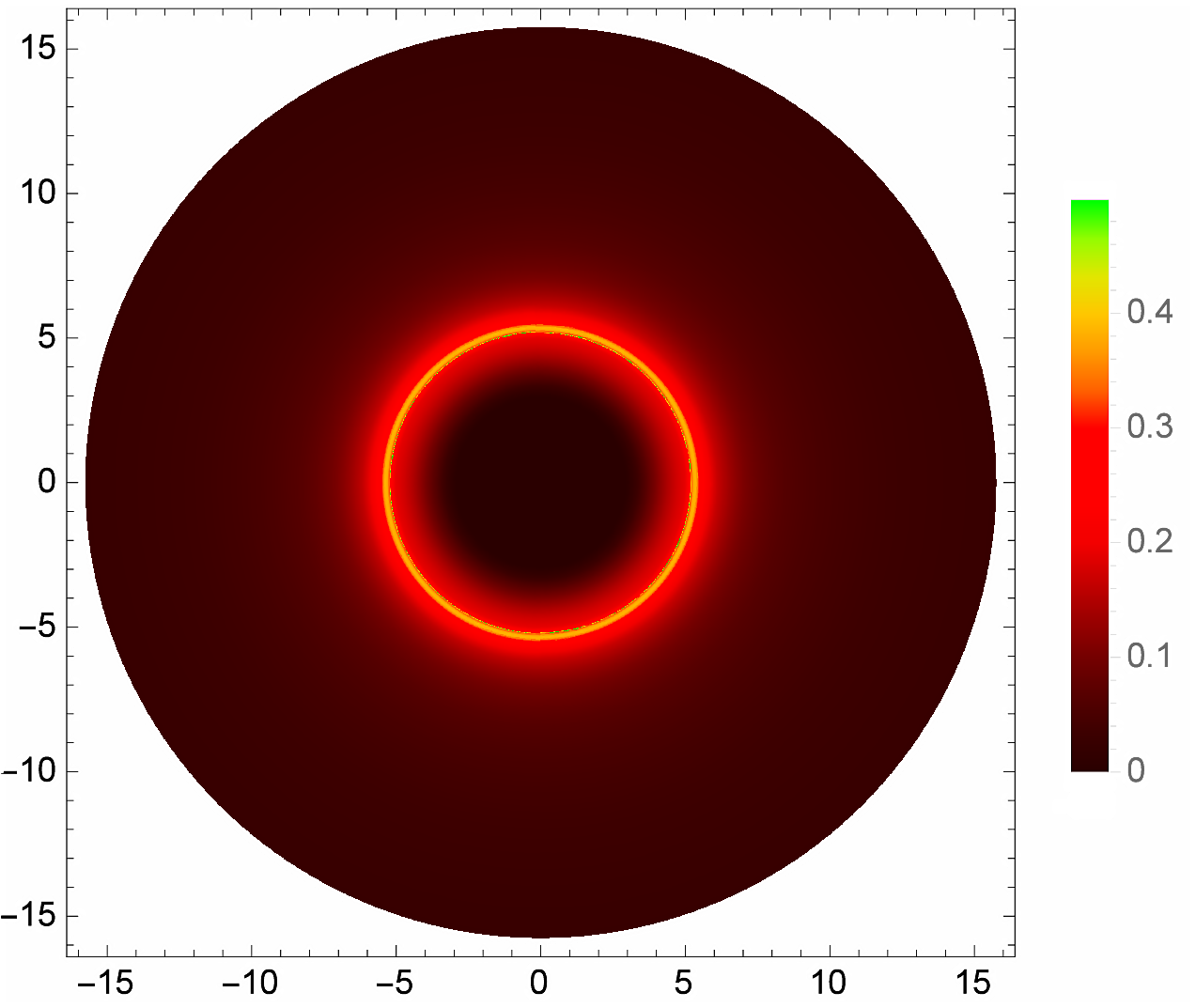}
\includegraphics[width=5.9cm,height=5.0cm]{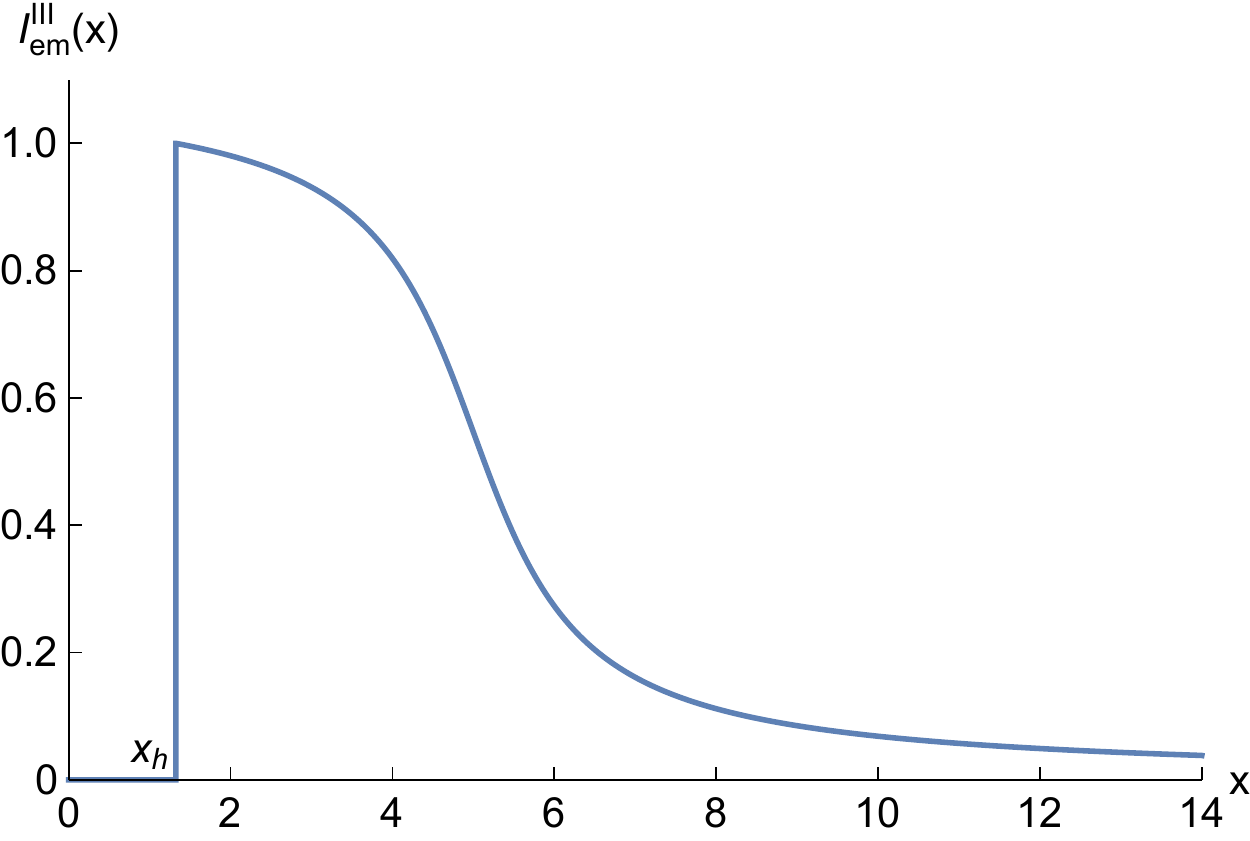}
\includegraphics[width=5.9cm,height=5.0cm]{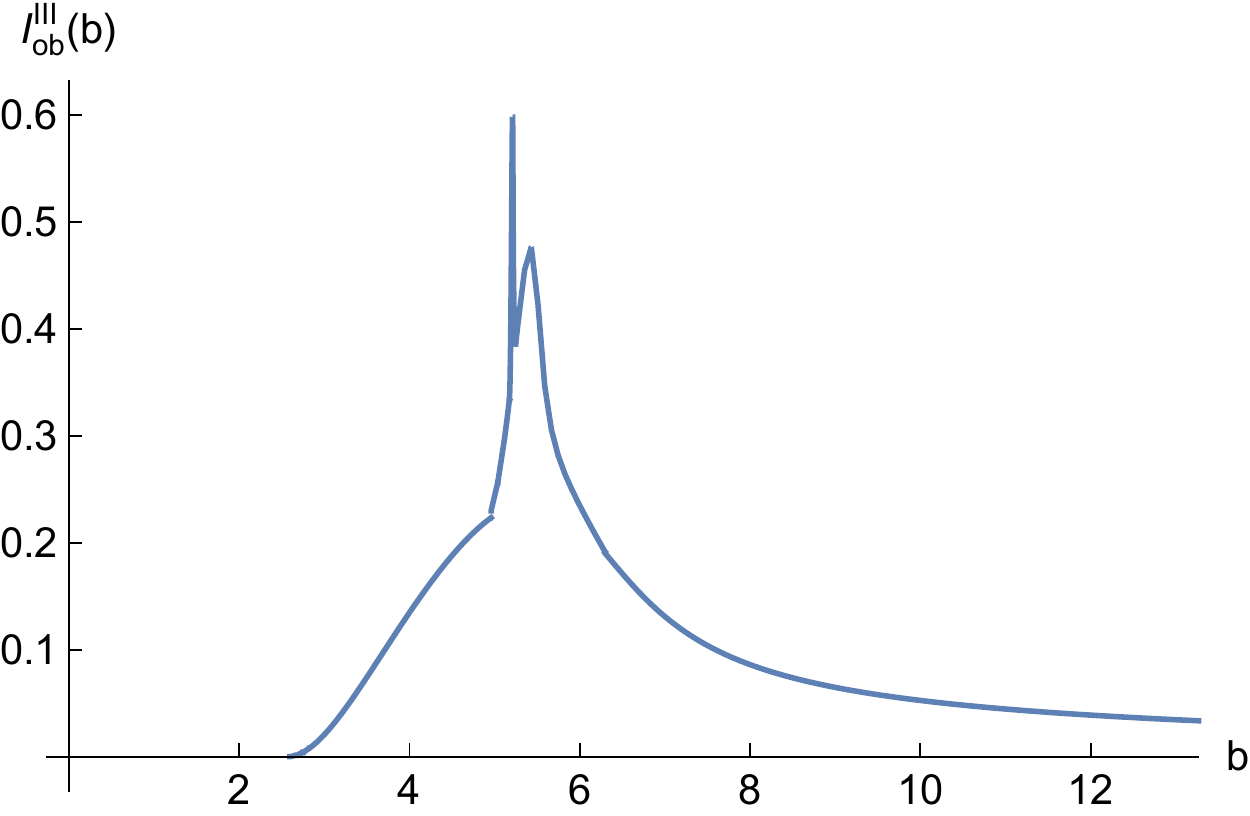}
\includegraphics[width=5.9cm,height=5.0cm]{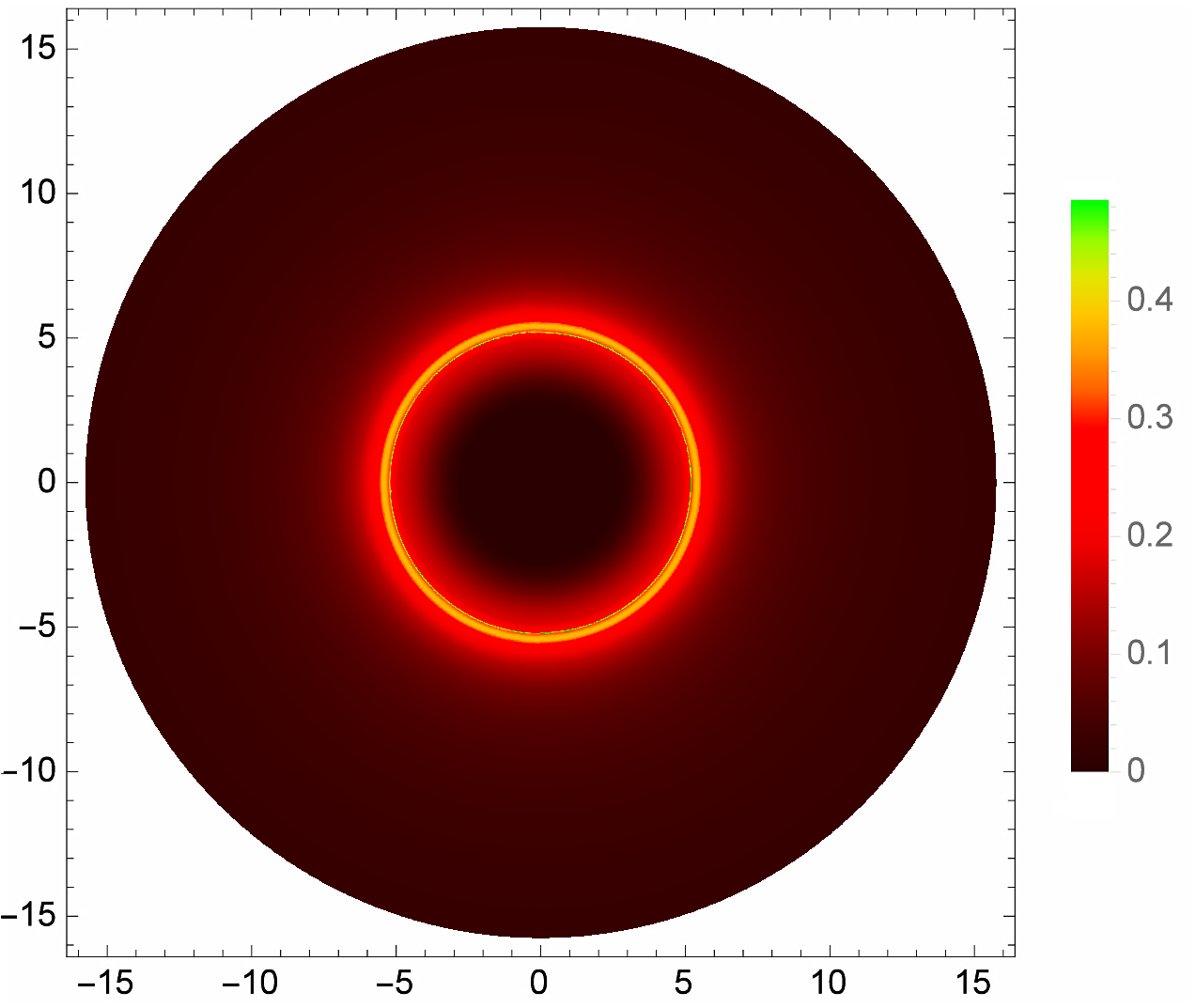}
\includegraphics[width=5.9cm,height=5.0cm]{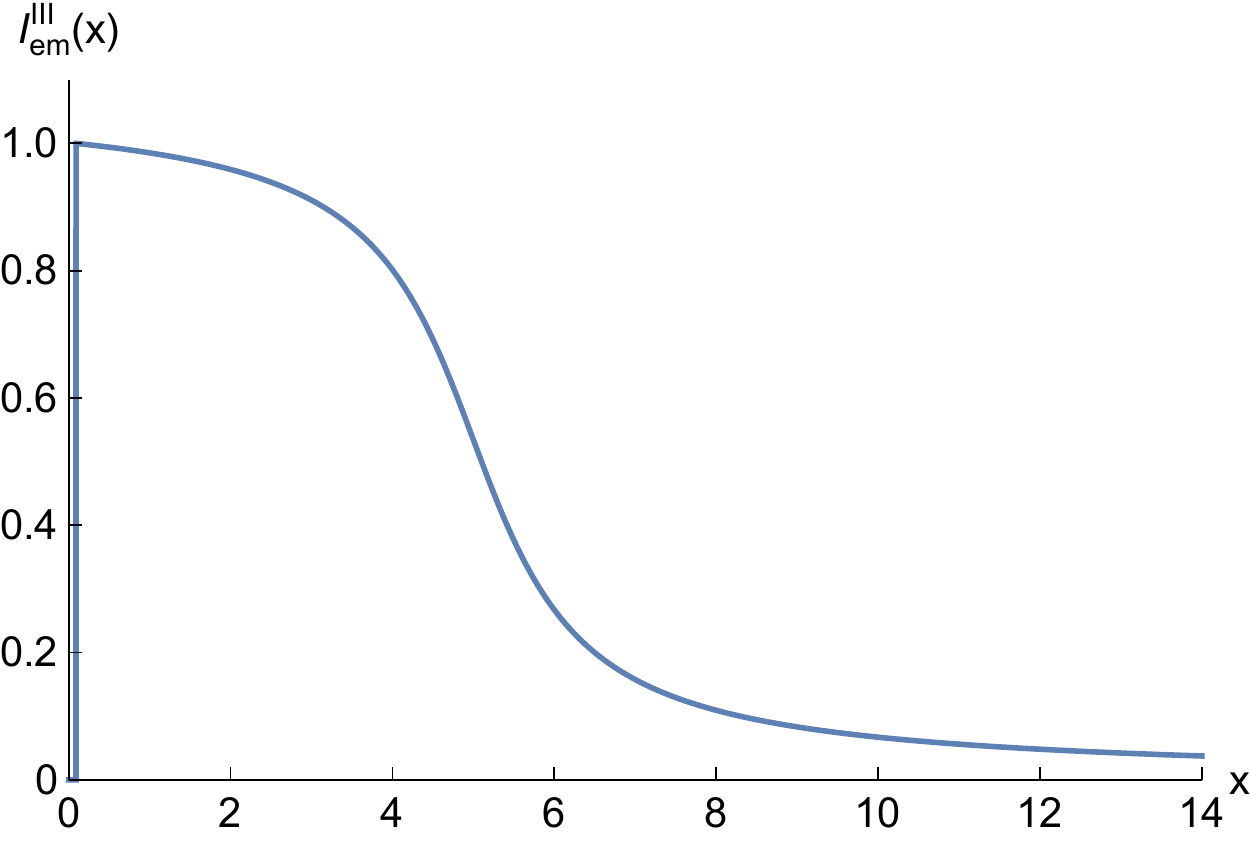}
\includegraphics[width=5.9cm,height=5.0cm]{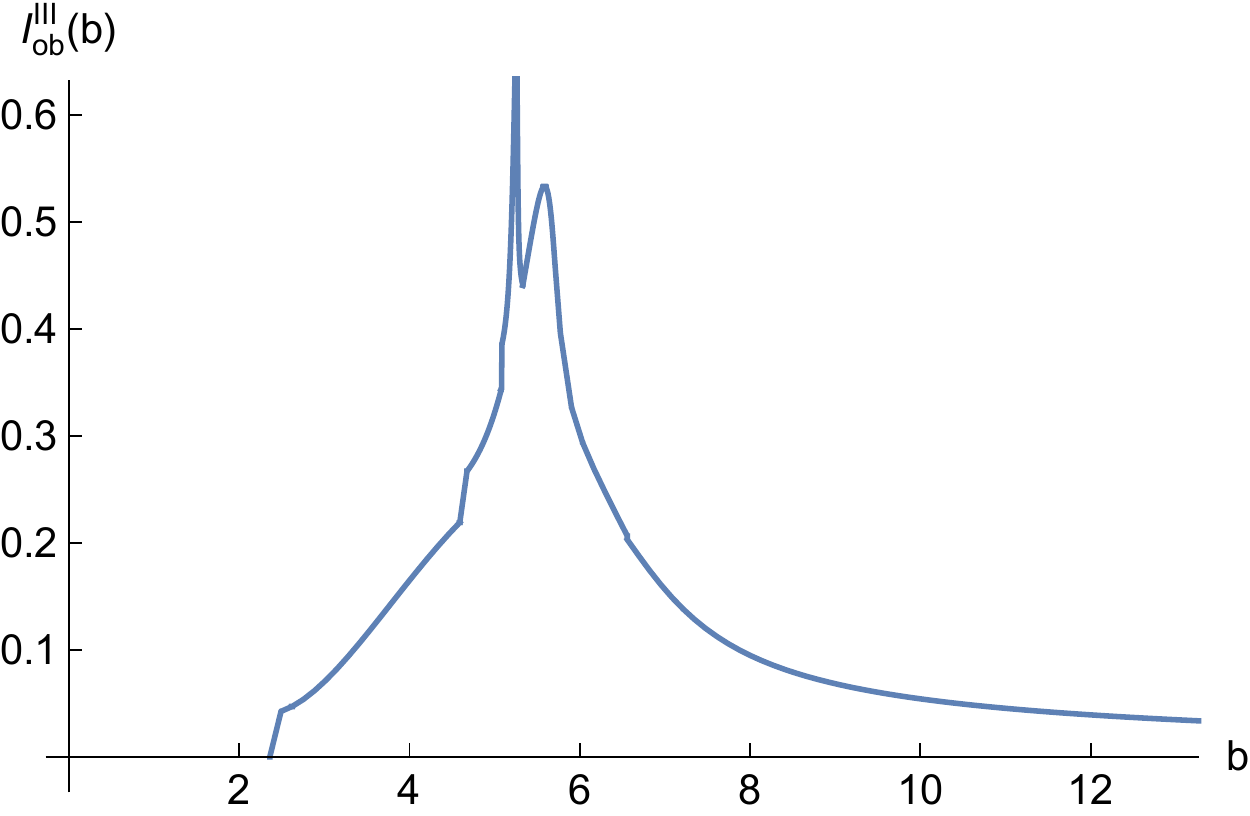}
\includegraphics[width=5.9cm,height=5.0cm]{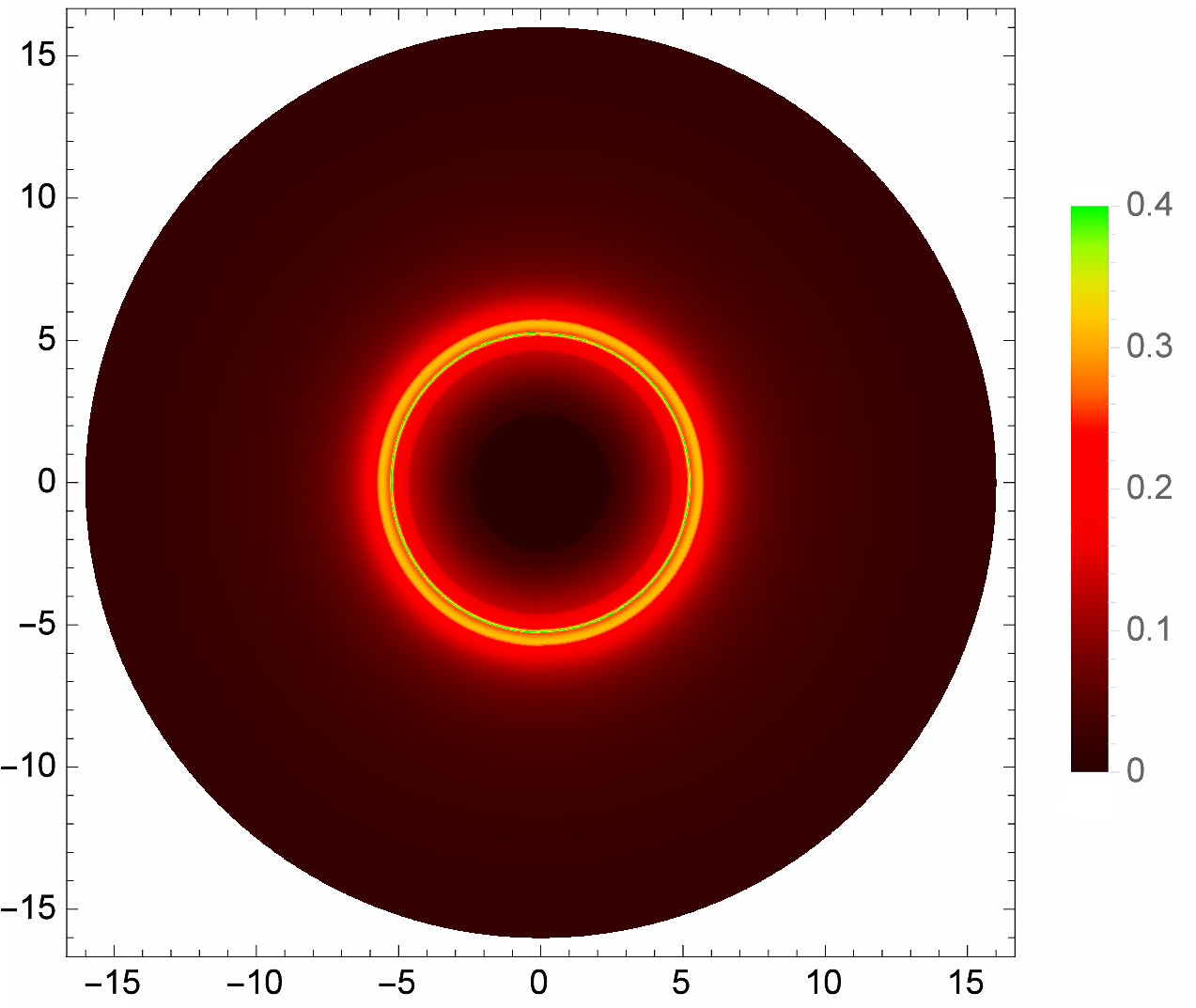}
\caption{The observational appearance of the BB solutions within accretion disk Model III with $a=0$ (Schwarzschild, top), $a=3/2$ (BH case, middle) and $a=5/2$ (WH case, bottom), viewed from a face-on orientation, and with a similar notation as in Fig. \ref{fig:model1}. In these plots $x_h=\sqrt{4M^2-a^2}$ is the horizon radius, since in Model III the emission goes all the way down to it. In the WH case (bottom panel) we have slightly displaced the beginning of this emission from $x=0$ (its throat) for numerical convergence reasons.}.
\label{fig:model3}
\end{figure*}

We can now proceed to study the optical appearance of the different families of BB solutions for the three models of emission above. To this end we depict in Figs. \ref{fig:model1}, \ref{fig:model2} and \ref{fig:model3} the emitted intensity (left), the observed intensity (middle), and the optical appearance (right) for each model of emission in the Schwarzschild case ($a=0$) and in the two samples of BB BH ($a=3/2$) and WH ($a=5/2$) solutions. As expected, the emission mode largely determines the qualitative shape of the optical appearance of the BB object.

In Model I, due to gravitational lensing in the observed intensity we clearly see the two isolated spikes representing the photon ring and lensing emissions, together with the more gradual decrease of the direct emission at larger impact parameter, neatly separated from each other. Therefore, the main contribution to the total luminosity in the optical appearance is provided by the direct emission yielding a wide ring, while inner to it we find the lensing ring and in the innermost region the barely visible to naked eye is the photon ring. In Model II, the direct, lensed, and photon ring emissions are overlapped in the observed intensity in a wider range of impact parameters. There are two peaks, one corresponding to the beginning of the direct emission which falls off until a superposition of the lensing and photon ring at almost coincident impact factor produces the large spike in this figure. However, the photon ring emission sharply falls off, quickly followed by the lensing one, until the direct emission dominates again. The net result is that in the optical appearance the lensing and photon rings are superimposed with the direct emission. The lensing ring contribution can be appreciated in this figure, though the one of the photon ring is highly diluted and barely visible. In Model III the direct observed region in impact parameter extends all the way down to the event horizon, increasing from there and getting again contributions at larger impact factors from the spike in the light ring first and in the lensing ring shortly after, before smoothly falling off to zero. The optical appearance in this case shows a narrow but somewhat brighter extended ring, made up of the contributions of the direct, lensed and photon ring emission, though as usual the latter can be safely ignored. This description of the optical appearances in these three models is completely consistent with the features obtained in similar images from the original description of the Schwarzschild black hole introduced in \cite{Gralla:2019xty}.

Moving forward to discussing the modifications of the BB solutions as compared to the Schwarzschild one ($a=0$), we first verify in Table \ref{table:II} that the contributions of both $I_{Lensed}$ and $I_{Photon}$ to the total luminosity as compared to the direct emission $I_{Direct}$ (including trajectories both above and below the critical impact parameter), though obviously emission-model-dependent, are significantly increased as $a$ grows. Indeed, when moving from $a=0$ (Schwarzschild) to $a=3/2$ (BB BH), the $I_{Lensed}$ contribution slightly rises for all the three models of the accretion disk, while those enhances are much more noticeable in the contributions from  $I_{Photon}$, though still pretty much negligible as compared to $I_{Direct}$. These increases are much more severe when moving to $a=5/2$ (WH branch), where the contribution of $I_{Lensed}$ can be twofold the original one (in Model II). Moreover, the contribution of the $I_{Photon}$ can be up to a factor  $\sim 5$ in Models II and III. This is due to the broadening of the impact factor region for both the lensed and photon trajectories in the WH case, as discussed in the ray-tracing of Sec. \ref{sec:III}, and that can be also seen in wider regions for the peaks of the observed intensities in the middle panels of Figs. \ref{fig:model1}, \ref{fig:model2} and \ref{fig:model3}.

Regarding the optical appearances (right panels), there are some tiny changes (for the BH case) but moderate ones (for the WH case) in the widths and intensities of the different light rings for all the emission models, which are barely visible in these plots for the BH case as compared to the Schwarzschild solution, but  much more noticeable in the WH one, as expected. This is particularly true for Model III, where the extended impact parameter region for both the lensing and photon rings (clearly visible in the corresponding observed luminosities) manifest as an additional boost of luminosity right in the middle of the direct emission, such that the combination of the lensing and photon ring contributions are now clearly visible. Large enhances of the contribution of the photon ring to the total luminosity have also been observed in certain models of compact objects with ``flattened" regions in the effective potential \cite{Gan:2021pwu}.

\section{Conclusion and prospects} \label{sec:V}

In this work we have considered the optical appearance (light rings and shadows) of an uniparametric family of (spherically symmetric) extensions of the Schwarzschild solution when illuminated by a thin accretion disk. Such a black bounce family smoothly interpolates between the Schwarzschild solution and two classes of solutions: regular black holes and traversable wormholes, and therefore it allows to compare the shadows cast by conceptually different objects on an equal footing. Moreover, this model has the additional advantage of having the last unstable orbit located exactly at the same critical impact parameter as in the original Schwarzschild solution for all the different BB configurations.

Using the ray-tracing procedure, we have classified the different light trajectories according to the number of orbits performed around the BB solution, splitting them into three main contributions according to the number of intersections with the equatorial plane: direct (one), lensed (two) and photon ring (three). Though in purity at the critical impact parameter the light ray would turn an infinite number of times, the subsequent contributions will be so demagnified that they can be safely ignored regarding their contributions to the total luminosity of the object. We found that, as the BB parameter increases, the impact parameter regions for these three contributions to the total luminosity are moderately enhanced, particularly in the WH case due to the retro-orbits flowing from the throat region thanks to the absence of an event horizon.

Next we considered a scenario of optically and geometrically thin accretion disks as the main source of illumination of the BB solutions, using three standard toy models whose emitted intensity peaks at the innermost stable circular orbit for time-like observers, at the last unstable circular orbit for photons, and near the horizon (in the BH case), or the throat (in the WH case). These three models are chosen  on the grounds that they simulate different physical scenarios and yield qualitative different observed emissions and their respective optical appearances for a given solution. The main modifications induced by the BB solutions as compared to the original Schwarzschild solution are an increase in the contributions of the lensed and photon ring emissions as compared to the direct one, which in the WH case are noticeable enough to be perceived at naked eye in some of the optical appearances plots.

The results found in this paper, though pointing to some differences in the shape of light rings and shadows of the BB solutions as compared to the Schwarzschild one, are in agreement with the running discussion on the community regarding the difficulty for testing hints of new Physics given the many elements involved in this analysis. In the model presented here, one would need to generalize it to include rotation \cite{Fran:2021pyi} in order to study the deviations in the circularity of the shadow when getting close to extremality rotation ratios. Moreover, the description of the accretion disks could be improved from the optically and geometrically thin modelling to a geometrically thick one, and the face-on orientation should be upgraded to consider modest inclinations of the disks and their effects in the optical appearances of the BB solution which, together with the addition of rotation, may significantly modify the total luminosity \cite{Beckwith:2004ae} and, therefore, the optical appearance. Finally, the presence of wormhole structures yields interesting new possibilities, such as shadows from objects without accretion disks due to contribution of those disks on the other side of the wormhole and flowing through the wormhole throat, or the generalization of our analysis to reflection-asymmetric wormholes since they produce effective potentials with two maxima (i.e. two unstable circular orbits), which can be capable to yield additional light rings \cite{Guerrero:2021pxt,Peng:2021osd}.

To conclude, whether the combination of all the above elements could be able to lead to further enhances in the brightness of the photon ring region, now including also additional contributions of the bands with $m>3$, is yet to be seen. Given the promises of the observational teams working on achieving better resolution for the optical appearance of black hole candidates in order to test GR to better precision, combined with some new ideas to test the photon sphere using interferometer \cite{John} or via correlated intensity fluctuations \cite{Hadar}, this field is ripe for the existing zoo of non-canonical compact objects to extract observational discriminators with respect to GR predictions.

\section*{Acknowledgements}
MG is funded by the predoctoral contract 2018-T1/TIC-10431 and acknowledges further support by the European Regional Development Fund under the Dora Plus scholarship grants.  DRG is funded by the \emph{Atracci\'on de Talento Investigador} programme of the Comunidad de Madrid (Spain) No. 2018-T1/TIC-10431, and acknowledges further support from the Ministerio de Ciencia, Innovaci\'on y Universidades (Spain) project No. PID2019-108485GB-I00/AEI/10.13039/501100011033, and the FCT projects No. PTDC/FIS-PAR/31938/2017 and PTDC/FIS-OUT/29048/2017. DS-CG is funded by the University of Valladolid (Spain), Ref. POSTDOC UVA20. This work is supported by the Spanish project  FIS2017-84440-C2-1-P (MINECO/FEDER, EU), the project PROMETEO/2020/079 (Generalitat Valenciana), and the Edital 006/2018 PRONEX (FAPESQ-PB/CNPQ, Brazil, Grant 0015/2019). This article is based upon work from COST Action CA18108, supported by COST (European Cooperation in Science and Technology). All images included in this paper where obtained with Mathematica@.


\begin{thebibliography}{100}

\bibitem{Akiyama:2019cqa}
K.~Akiyama \textit{et al.} [Event Horizon Telescope],
%``First M87 Event Horizon Telescope Results. I. The Shadow of the Supermassive Black Hole,''
Astrophys. J. Lett. \textbf{875} (2019) L1.

\bibitem{Falcke:1999pj}
H.~Falcke, F.~Melia and E.~Agol,
%``Viewing the shadow of the black hole at the galactic center,''
Astrophys. J. Lett. \textbf{528} (2000) L13.

\bibitem{Perlick:2021aok}
V.~Perlick and O.~Y.~Tsupko,
%``Calculating black hole shadows: review of analytical studies,''
[arXiv:2105.07101 [gr-qc]].

\bibitem{Cardoso:2019rvt}
V.~Cardoso and P.~Pani,
%``Testing the nature of dark compact objects: a status report,''
Living Rev. Rel. \textbf{22} (2019) 4.

\bibitem{Psaltis:2020lvx}
D.~Psaltis \textit{et al.} [Event Horizon Telescope],
%``Gravitational Test Beyond the First Post-Newtonian Order with the Shadow of the M87 Black Hole,''
Phys. Rev. Lett. \textbf{125} (2020) 141104.

\bibitem{Bardeen}
J. M. Bardeen, in Black Holes (Les Astres Occlus), C. De Witt and B. S. de Witt (Eds.), Gordon\&Breach, New York, 1973.

\bibitem{Bozza:2002zj}
V.~Bozza,
%``Gravitational lensing in the strong field limit,''
Phys. Rev. D \textbf{66} (2002) 103001.

\bibitem{Johannsen:2010ru}
T.~Johannsen and D.~Psaltis,
%``Testing the No-Hair Theorem with Observations in the Electromagnetic Spectrum: II. Black-Hole Images,''
Astrophys. J. \textbf{718} (2010) 446.

\bibitem{Atamurotov:2013sca}
F.~Atamurotov, A.~Abdujabbarov and B.~Ahmedov,
%``Shadow of rotating non-Kerr black hole,''
Phys. Rev. D \textbf{88} (2013) 064004.

\bibitem{Cunha:2015yba}
P.~V.~P.~Cunha, C.~A.~R.~Herdeiro, E.~Radu and H.~F.~Runarsson,
%``Shadows of Kerr black holes with scalar hair,''
Phys. Rev. Lett. \textbf{115} (2015)  211102.

\bibitem{Abdujabbarov:2016hnw}
A.~Abdujabbarov, M.~Amir, B.~Ahmedov and S.~G.~Ghosh,
%``Shadow of rotating regular black holes,''
Phys. Rev. D \textbf{93} (2016) 104004.

\bibitem{Held:2019xde}
A.~Held, R.~Gold and A.~Eichhorn,
%``Asymptotic safety casts its shadow,''
JCAP \textbf{06} (2019) 029.

\bibitem{Kumar:2020owy}
R.~Kumar and S.~G.~Ghosh,
%``Rotating black holes in $4D$ Einstein-Gauss-Bonnet gravity and its shadow,''
JCAP \textbf{07} (2020) 053.

\bibitem{Xavier:2020egv}
S.~V.~M.~C.~B.~Xavier, P.~V.~P.~Cunha, L.~C.~B.~Crispino and C.~A.~R.~Herdeiro,
%``Shadows of charged rotating black holes: Kerr\textendash{}Newman versus Kerr\textendash{}Sen,''
Int. J. Mod. Phys. D \textbf{29} (2020) 2041005.

\bibitem{Wei:2020ght}
S.~W.~Wei and Y.~X.~Liu,
%``Testing the nature of Gauss-Bonnet gravity by four-dimensional rotating black hole shadow,''
Eur. Phys. J. Plus \textbf{136} (2021)  436.

\bibitem{Herdeiro:2021lwl}
C.~A.~R.~Herdeiro, A.~M.~Pombo, E.~Radu, P.~V.~P.~Cunha and N.~Sanchis-Gual,
%``The imitation game: Proca stars that can mimic the Schwarzschild shadow,''
JCAP \textbf{04} (2021) 051.

\bibitem{Devi:2021ctm}
S.~Devi, S.~Chakrabarti and B.~R.~Majhi,
%``Shadow of quantum extended Kruskal black hole and its super-radiance property,''
[arXiv:2105.11847 [gr-qc]].

\bibitem{Hou:2021okc}
Y.~Hou, M.~Guo and B.~Chen,
%``Revisiting the shadow of braneworld black holes,''
[arXiv:2103.04369 [gr-qc]].

\bibitem{Narayan:2019imo}
R.~Narayan, M.~D.~Johnson and C.~F.~Gammie,
%``The Shadow of a Spherically Accreting Black Hole,''
Astrophys. J. Lett. \textbf{885} (2019) L33.

\bibitem{Cunha:2019hzj}
P.~Cunha, V.P., N.~A.~Eir\'o, C.~A.~R.~Herdeiro and J.~P.~S.~Lemos,
%``Lensing and shadow of a black hole surrounded by a heavy accretion disk,''
JCAP \textbf{03} (2020) 035.

\bibitem{Boero:2021afh}
E.~F.~Boero and O.~M.~Moreschi,
%``Strong gravitational lens image of the M87 black hole with a simple accreting matter model,''
[arXiv:2105.07075 [gr-qc]].


\bibitem{Gralla:2019xty}
S.~E.~Gralla, D.~E.~Holz and R.~M.~Wald,
Phys. Rev. D \textbf{100} (2019)  024018.

\bibitem{Glampedakis:2021oie}
K.~Glampedakis and G.~Pappas,
%``Can supermassive black hole shadows test the Kerr metric?,''
[arXiv:2102.13573 [gr-qc]].

\bibitem{Junior:2021atr}
H.~C.~D.~Lima, Junior., L.~C.~B.~Crispino, P.~V.~P.~Cunha and C.~A.~R.~Herdeiro,
%``Can different black holes cast the same shadow?,''
Phys. Rev. D \textbf{103} (2021)  084040.

\bibitem{Chael:2021rjo}
A.~Chael, M.~D.~Johnson and A.~Lupsasca,
%``Observing the Inner Shadow of a Black Hole: A Direct View of the Event Horizon,''
[arXiv:2106.00683 [astro-ph.HE]].


\bibitem{Zeng:2020dco}
X.~X.~Zeng, H.~Q.~Zhang and H.~Zhang,
%``Shadows and photon spheres with spherical accretions in the four-dimensional Gauss\textendash{}Bonnet black hole,''
Eur. Phys. J. C \textbf{80} (2020)  872.

\bibitem{Zeng:2020vsj}
X.~X.~Zeng and H.~Q.~Zhang,
%``Influence of quintessence dark energy on the shadow of black hole,''
Eur. Phys. J. C \textbf{80} (2020)  1058.

\bibitem{Qin:2020xzu}
X.~Qin, S.~Chen and J.~Jing,
%``Image of a regular phantom compact object and its luminosity under spherical accretions,''
Class. Quant. Grav. \textbf{38} (2021)  115008.

\bibitem{Lima:2020auu}
H.~C.~D.~Lima, C.~L.~Benone and L.~C.~B.~Crispino,
%``Scalar absorption: Black holes versus wormholes,''
Phys. Rev. D \textbf{101} (2020) 124009.

\bibitem{He:2021htq}
K.~J.~He, S.~Guo, S.~C.~Tan and G.~P.~Li,
%``The feature of shadow images and observed luminosity of the Bardeen black hole surrounded by different accretions,''
[arXiv:2103.13664 [hep-th]].


\bibitem{Peng:2020wun}
J.~Peng, M.~Guo and X.~H.~Feng,
%``Influence of Quantum Correction on the Black Hole Shadows, Photon Rings and Lensing Rings,''
[arXiv:2008.00657 [gr-qc]].

\bibitem{Eichhorn:2021iwq}
A.~Eichhorn and A.~Held,
%``From a locality-principle for new physics to image features of regular spinning black holes with disks,''
JCAP \textbf{05} (2021) 073.

\bibitem{Eichhorn:2021etc}
A.~Eichhorn and A.~Held,
%``Image features of spinning regular black holes based on a locality principle,''
[arXiv:2103.07473 [gr-qc]].


\bibitem{Gan:2021pwu}
Q.~Gan, P.~Wang, H.~Wu and H.~Yang,
%``Photon Spheres and Spherical Accretion Image of a Hairy Black Hole,''
[arXiv:2104.08703 [gr-qc]].

\bibitem{Li:2021riw}
G.~P.~Li and K.~J.~He,
%``Shadows and rings of the Kehagias-Sfetsos black hole surrounded by thin disk accretion,''
[arXiv:2105.08521 [gr-qc]].

\bibitem{1865010}
Q.~Gan, P.~Wang, H.~Wu and H.~Yang,
%``Photon Ring and Observational Appearance of a Hairy Black Hole,''
[arXiv:2105.11770 [gr-qc]].

\bibitem{Shaikh:2021cvl}
R.~Shaikh, S.~Paul, P.~Banerjee and T.~Sarkar,
%``Shadows and thin accretion disk images of the $\gamma$-metric,''
[arXiv:2105.12057 [gr-qc]].

\bibitem{Psaltis:2018xkc}
D.~Psaltis,
%``Testing General Relativity with the Event Horizon Telescope,''
Gen. Rel. Grav. \textbf{51} (2019) 137.

\bibitem{Simpson:2018tsi}
A.~Simpson and M.~Visser,
%``Black-bounce to traversable wormhole,''
JCAP \textbf{02} (2019) 042.

\bibitem{Churilova:2019cyt}
M.~S.~Churilova and Z.~Stuchlik,
%``Ringing of the regular black-hole/wormhole transition,''
Class. Quant. Grav. \textbf{37} (2020)  075014.

\bibitem{Lobo:2020kxn}
F.~S.~N.~Lobo, A.~Simpson and M.~Visser,
%``Dynamic thin-shell black-bounce traversable wormholes,''
Phys. Rev. D \textbf{101} (2020)  124035.

\bibitem{Huang:2019arj}
H.~Huang and J.~Yang,
%``Charged Ellis Wormhole and Black Bounce,''
Phys. Rev. D \textbf{100} (2019)  124063.

\bibitem{Nascimento:2020ime}
J.~R.~Nascimento, A.~Y.~Petrov, P.~J.~Porfirio and A.~R.~Soares,
%``Gravitational lensing in black-bounce spacetimes,''
Phys. Rev. D \textbf{102} (2020)  044021.

\bibitem{Lobo:2020ffi}
F.~S.~N.~Lobo, M.~E.~Rodrigues, M.~V.~d.~S.~Silva, A.~Simpson and M.~Visser,
%``Novel black-bounce spacetimes: wormholes, regularity, energy conditions, and causal structure,''
Phys. Rev. D \textbf{103} (2021)  084052.

\bibitem{Tsukamoto:2020bjm}
N.~Tsukamoto,
%``Gravitational lensing in the Simpson-Visser black-bounce spacetime in a strong deflection limit,''
Phys. Rev. D \textbf{103} (2021) 024033.

\bibitem{Zhou:2020zys}
T.~Y.~Zhou and Y.~Xie,
%``Precessing and periodic motions around a black-bounce/traversable wormhole,''
Eur. Phys. J. C \textbf{80} (2020)  1070.

\bibitem{Mazza:2021rgq}
J.~Mazza, E.~Franzin and S.~Liberati,
%``A novel family of rotating black hole mimickers,''
JCAP \textbf{04} (2021) 082.

\bibitem{Shaikh:2021yux}
R.~Shaikh, K.~Pal, K.~Pal and T.~Sarkar,
%``Constraining alternatives to the Kerr black hole,''
[arXiv:2102.04299 [gr-qc]].

\bibitem{Cheng:2021hoc}
X.~T.~Cheng and Y.~Xie,
%``Probing a black-bounce, traversable wormhole with weak deflection gravitational lensing,''
Phys. Rev. D \textbf{103} (2021) 064040.

\bibitem{Islam:2021ful}
S.~U.~Islam, J.~Kumar and S.~G.~Ghosh,
%``Strong gravitational lensing by rotating Simpson--Visser black holes,''
[arXiv:2104.00696 [gr-qc]].

\bibitem{Fran:2021pyi}
E.~Franzin, S.~Liberati, J.~Mazza, A.~Simpson and M.~Visser,
%``Charged black-bounce spacetimes,''
[arXiv:2104.11376 [gr-qc]].


\bibitem{Bronnikov:2021liv}
K.~A.~Bronnikov, R.~A.~Konoplya and T.~D.~Pappas,
%``General parametrization of wormhole spacetimes and its application to shadows and quasinormal modes,''
Phys. Rev. D \textbf{103} (2021) 124062.


\bibitem{Tsukamoto:2021caq}
N.~Tsukamoto,
%``Gravitational lensing by two photon spheres in a black-bounce spacetime in strong deflection limits,''
[arXiv:2105.14336 [gr-qc]].

\bibitem{Zeng:2021dlj}
X.~X.~Zeng, G.~P.~Li and K.~J.~He,
%``The shadows and observational appearance of a noncommutative black hole surrounded by various profiles of accretions,''
[arXiv:2106.14478 [hep-th]].


\bibitem{Poisson:1989zz}
E.~Poisson and W.~Israel,
%``Inner-horizon instability and mass inflation in black holes,''
Phys. Rev. Lett. \textbf{63} (1989) 1663.

\bibitem{Olmo:2013gqa}
G.~J.~Olmo, D.~Rubiera-Garcia and H.~Sanchis-Alepuz,
%``Geonic black holes and remnants in Eddington-inspired Born-Infeld gravity,''
Eur. Phys. J. C \textbf{74} (2014) 2804.

\bibitem{Olmo:2015bya}
G.~J.~Olmo, D.~Rubiera-Garcia and A.~Sanchez-Puente,
%``Geodesic completeness in a wormhole spacetime with horizons,''
Phys. Rev. D \textbf{92} (2015) 044047.

\bibitem{Bejarano:2017fgz}
C.~Bejarano, G.~J.~Olmo and D.~Rubiera-Garcia,
%``What is a singular black hole beyond General Relativity?,''
Phys. Rev. D \textbf{95} (2017) 064043.

\bibitem{VisserBook}
M. Visser, {\it ``Lorentzian Wormholes"} (Springer-Verlag, NY, 1996).


\bibitem{Carballo-Rubio:2019fnb}
R.~Carballo-Rubio, F.~Di Filippo, S.~Liberati and M.~Visser,
%``Geodesically complete black holes,''
Phys. Rev. D \textbf{101} (2020) 084047.


\bibitem{Page:1974he}
D.~N.~Page and K.~S.~Thorne,
%``Disk-Accretion onto a Black Hole. Time-Averaged Structure of Accretion Disk,''
Astrophys. J. \textbf{191} (1974) 499.

\bibitem{Riaz:2019bkv}
S.~Riaz, D.~Ayzenberg, C.~Bambi and S.~Nampalliwar,
%``Reflection spectra of thick accretion discs,''
Mon. Not. Roy. Astron. Soc. \textbf{491} (2020) 417.


\bibitem{Beckwith:2004ae}
K.~Beckwith and C.~Done,
%``Extreme gravitational lensing near rotating black holes,''
Mon. Not. Roy. Astron. Soc. \textbf{359} (2005) 1217.

\bibitem{Guerrero:2021pxt}
M.~Guerrero, G.~J.~Olmo and D.~Rubiera-Garcia,
%``Double shadows of reflection-asymmetric wormholes supported by positive energy thin-shells,''
JCAP \textbf{04} (2021) 066.

\bibitem{Peng:2021osd}
J.~Peng, M.~Guo and X.~H.~Feng,
%``Observational Signature and Additional Photon Rings of Asymmetric Thin-shell Wormhole,''
[arXiv:2102.05488 [gr-qc]].


\bibitem{John}
M. D. Johnson, et al. Sci. Adv. \textbf{6} (2020) eaaz1310.

\bibitem{Hadar}S. Hadar, M. D. Johnson, A. Lupsasca and G. N. Wong, Phys. Rev. D \textbf{103} (2021) 104038.

	
\end{thebibliography}
\end{document}